\documentclass[conference]{IEEEtran}
\usepackage{cite}
\usepackage{amsmath,amssymb,amsfonts}
\usepackage{algorithmic}
\usepackage{graphicx}
\usepackage{textcomp}
\usepackage{xcolor}
\let\labelindent\relax
\usepackage{enumitem}

\usepackage{dsfont}
\usepackage{bm}    
\usepackage{booktabs}
\usepackage{comment}
\usepackage{hyperref}
\usepackage{multirow}
\usepackage{subfigure}
\usepackage{soul}
\usepackage{tikz}
\usepackage{xspace}
\usepackage[flushleft]{threeparttable}
\usepackage{caption}
\def\eg{\emph{e.g.,}\xspace}

\def\ie{\emph{i.e.,}\xspace}

\newcommand{\pie}[1]{%
\begin{tikzpicture}
 \draw (0,0) circle (0.75ex);\fill (0.75ex,0) arc (0:(-#1+90):0.75ex) -- (0,0) -- cycle;
 \fill (0.75ex,0) arc (0:(#1-90):0.75ex) -- (0,0) -- cycle;
\end{tikzpicture}%
}

\usepackage{authblk}

\begin{document}

\title{ The ``Beatrix'' Resurrections:\\ Robust Backdoor Detection via Gram Matrices}

\author[$\dagger$]{Wanlun Ma}
\author[$\ddagger$]{Derui Wang}
\author[$\ddagger$]{Ruoxi Sun}
\author[$\ddagger$]{Minhui Xue}
\author[$\dagger$]{Sheng Wen}
\author[$\dagger$]{Yang Xiang}
\affil[$\dagger$]{Swinburne University of Technology, Australia}
\affil[$\ddagger$]{CSIRO’s Data61, Australia}

\IEEEoverridecommandlockouts
\makeatletter\def\@IEEEpubidpullup{6.5\baselineskip}\makeatother
\IEEEpubid{\parbox{\columnwidth}{
    Network and Distributed System Security (NDSS) Symposium 2023\\
    28 February - 4 March 2023, San Diego, CA, USA\\
    ISBN 1-891562-83-5\\
    https://dx.doi.org/10.14722/ndss.2023.23069\\
    www.ndss-symposium.org
}
\hspace{\columnsep}\makebox[\columnwidth]{}}

\maketitle

\begin{abstract}
Deep Neural Networks (DNNs) are susceptible to backdoor attacks during training. The model corrupted in this way functions normally, but when triggered by certain patterns in the input, produces a predefined target label. 
Existing defenses usually rely on the assumption of the universal backdoor setting in which poisoned samples share the same uniform trigger. However, recent advanced backdoor attacks show that this assumption is no longer valid in dynamic backdoors where the triggers vary from input to input, thereby defeating the existing defenses.

In this work, we propose a novel technique, \textit{Beatrix} (\underline{b}ackdoor d\underline{e}tection via Gr\underline{a}m ma\underline{trix}).
Beatrix utilizes Gram matrix to capture not only the feature correlations but also the appropriately high-order information of the representations. By learning class-conditional statistics from activation patterns of normal samples, Beatrix can identify poisoned samples by capturing the anomalies in activation patterns. 
To further improve the performance in identifying target labels, Beatrix leverages kernel-based testing without making any prior assumptions on representation distribution. 
We demonstrate the effectiveness of our method through theoretical justifications and extensive comparisons with state-of-the-art defensive techniques. 
The experimental results show that our approach achieves an F1 score of 91.1\% in detecting dynamic backdoors, while the state of the art can only reach 36.9\%.
\end{abstract}


\section{Introduction}

\begin{table*}[t]
\centering
\caption{\label{table: summary of existing defenses} A summary of the existing defenses and our work.}
\resizebox{0.92\textwidth}{!}{
\begin{threeparttable}
\begin{tabular}{ccccccccccc} 
\toprule
\multirow{2}{*}{\textbf{Type}}                                                            & \multirow{2}{*}{\textbf{Approaches}} & \multicolumn{3}{c}{\textbf{Detection Target}}       & \multirow{2}{*}{\begin{tabular}[c]{@{}c@{}}\textbf{Black-box}\\~\textbf{Access}\end{tabular}} & \multirow{2}{*}{\begin{tabular}[c]{@{}c@{}}\textbf{No Need of}\\\textbf{Clean Data}\end{tabular}} & \multirow{2}{*}{\begin{tabular}[c]{@{}c@{}}\textbf{All-to-all}\\\textbf{Attack}\end{tabular}} & \multicolumn{3}{c}{\textbf{Trigger Assumption}}  \\ 
\cmidrule{3-5}\cmidrule{9-11}
& & \multicolumn{1}{c}{\textbf{Input}} & \textbf{Model} & \textbf{Trigger} & & & & \textbf{Universal} & \textbf{Partial} & \textbf{Dynamic} \\ 
\midrule
\multirow{3}{*}{\begin{tabular}[c]{@{}c@{}}I:\ Input \\Masking\end{tabular}} & STRIP~\cite{gao2019strip} & \pie{360} & \pie{90} & \pie{90} & \pie{360} & \pie{90} & \pie{90} & \pie{360} & \pie{90} & \pie{90}\\ 
& Februus~\cite{doan2020februus}  & \pie{360} & \pie{90} & \pie{360} & \pie{90} & \pie{90} & \pie{360} & \pie{360} & \pie{90} & \pie{90}  \\ 
& SentiNet~\cite{chou2020sentinet} & \pie{360} & \pie{90} & \pie{360} & \pie{90} & \pie{90} & \pie{360} & \pie{360} & \pie{90} & \pie{90}\\ 
\midrule
\multirow{3}{*}{\begin{tabular}[c]{@{}c@{}}II:\ Model \\Inspection\end{tabular}} & Neural Cleanse~\cite{wang2019neural} & \pie{90} & \pie{360} & \pie{360} & \pie{90} & \pie{90} & \pie{90} & \pie{360} & \pie{90} & \pie{90} \\ 
& ABS~\cite{liu2019abs} & \pie{90} & \pie{360} & \pie{360} & \pie{90} & \pie{90} & \pie{90} & \pie{360} & \pie{90} & \pie{90} \\ 
& MNTD~\cite{xu2021detecting}  & \pie{90} & \pie{360} & \pie{90} & \pie{360} & \pie{90} & \pie{360} & \pie{360} & \pie{90} & \pie{90} \\ 
\midrule
\multirow{5}{*}{\begin{tabular}[c]{@{}c@{}}III:\ Feature\\
Representation\end{tabular}} & Activation-Clustering~\cite{chen2019detecting} & \pie{90} & \pie{360} & \pie{90} & \pie{90} & \pie{360} & \pie{360} & \pie{360} & \pie{90} & \pie{90} \\ 
& Spectral-Signature~\cite{tran2018spectral} & \pie{90} & \pie{360} & \pie{90} & \pie{90} & \pie{360} & \pie{360} & \pie{360} & \pie{90} & \pie{90} \\ 
& SPECTRE~\cite{hayase2021spectre} & \pie{90} & \pie{360} & \pie{90} & \pie{90} & \pie{90} & \pie{360} & \pie{360} & \pie{90} & \pie{90} \\ 
& SCAn~\cite{tang2021demon} & \pie{360} & \pie{360} & \pie{90} & \pie{90} & \pie{90} & \pie{360} & \pie{360} & \pie{360} & \pie{90}  \\ 
& \textbf{Beatrix (our work)} & \pie{360} & \pie{360} & \pie{90} & \pie{90} & \pie{90} & \pie{360} & \pie{360} & \pie{360} & \pie{360} \\
\bottomrule
\end{tabular}
\begin{tablenotes}
\item[]\footnotesize\pie{90}: the item is not supported by the defense; \pie{360}: the item is supported by the defense.
\end{tablenotes}
\end{threeparttable}
}
\vspace{-3mm}
\end{table*}

With an explosive growth of Machine Learning (ML) and Artificial Intelligence (AI), Deep Neural Networks (DNNs) are widely adopted in many significant real-world and security-critical scenarios, including facial recognition~\cite{sharif2016accessorize}, self-driving navigation~\cite{sitawarin2018darts}, and medical diagnosis~\cite{rajkomar2018scalable}. Despite these surprising advances, it has been known that DNNs suffer from severe security issues, such as privacy leakage~\cite{shokri2017membership}, adversarial attacks~\cite{szegedy2013intriguing, goodfellow2014explaining} and backdoor attacks (a.k.a. Trojan attacks)~\cite{gu2019badnets, chen2017targeted}. In particular, the backdoor attack is a technique of embedding a hidden malicious functionality into the DNN, which is activated only when a certain trigger appears. 
This hidden functionality is usually the misclassification of an input sample to the attacker's desired target class, given the presence of a predefined trigger.
For example, a stop sign corrupted by a few pieces of tape will be recognized as a speed limit sign by the navigation system in self-driving cars, which may lead to fatal consequences~\cite{eykholt2018robust}.

BadNets~\cite{gu2019badnets} is one of the first works to study the threat of neural backdoors. After that, many variants of backdoor attacks have been proposed~\cite{chen2017targeted, liu2018trojaning, yao2019latent, jia2022badencoder}. Despite varying in mechanisms and scenarios, all these existing backdoor attacks are premised on adopting a universal (or sample-agnostic) trigger, \ie different poisoned samples carry the same trigger. This uniform backdoor trigger becomes the Achilles' heel of the backdoor attacks. Based on the fact that the trigger is fixed and universal, existing defensive techniques~\cite{wang2019neural, liu2019abs, gao2019strip, chou2020sentinet, chen2019detecting} can easily reconstruct or detect the trigger according to the same behaviors among different poisoned samples. 
For example, Neural Cleanse~\cite{wang2019neural} utilizes an optimization scheme to synthesize potential trigger patterns that can convert all benign images of other classes to a specific class. The synthesized trigger pattern with abnormally small norm is considered as the attack pattern used by the adversary. Additionally, a defender can also perform run-time detection on each input sample. To examine a malicious input, STRIP~\cite{gao2019strip} superimposes an input image to a set of randomly selected images and measures the entropy of the prediction outputs. If the predictions of the blending images are consistent (\ie low entropy of prediction outputs), this input is regarded as a malicious one. In addition, SentiNet~\cite{chou2020sentinet} exploits the explanation technique (\eg Grad-CAM~\cite{selvaraju2017grad}) to locate a potential trigger region by finding a highly salient contiguous region of a given input.

Witness to the success of the existing defenses, one might think that the threat of backdoor attacks is mitigated or neutralized.
Unfortunately, the crucial weakness of such static and sample-agnostic trigger became known to adversaries and they started exploring more advanced approaches in their attacks. In the new attack paradigms, backdoor triggers (referred to as \textit{dynamic}~\cite{nguyen2020input} or \textit{sample-specific}~\cite{li2021invisible} triggers) vary from sample to sample. 
The success of existing defensive techniques~\cite{wang2019neural, gao2019strip, chou2020sentinet} mostly relies on the assumption that the triggers are sample-agnostic. 
However, the sample-specific backdoor attacks break the fundamental assumption of the existing defensive techniques, as the dynamic backdoor introduces diverse information into the trigger pattern, which makes it harder for the defender to model the trigger. 
As shown in Table~\ref{table: summary of existing defenses}, current backdoor defensive techniques mainly focus on universal backdoor attacks, leaving dynamic backdoor attacks as an unaddressed crucial threat to DNNs (see more discussions in Section~\ref{sec: Limitations against Dynamic Backdoors}).
Although a poisoned sample is misclassified to a target label, its intermediate representation has been shown to be different from those of the normal samples in the target class~\cite{tran2018spectral, chen2019detecting, tang2021demon}. 
This observation provides an important indicator to distinguish the malicious samples from the normal ones. 
However, when you zoom in with order information, out-of-distribution (OOD) samples present more details than trojaned ones since OOD detection requires much higher-order in screening out OOD samples~\cite{liang2017enhancing, lee2018simple, zisselman2020deep, sastry2020detecting}. This observation renders OOD detection methods to be less appealing to trojan detection as the OOD detection signal is too strong. 
To further demystify the reasons, OOD detection requires sufficient and uncontaminated data as \textit{a priori} knowledge, which is not practical in backdoor detection. Moreover, although the Gram matrix based OOD detector achieves successful performance~\cite{sastry2020detecting}, our experimental results demonstrate that it lacks robustness (using fragile deviation metrics) and efficiency (computing an over-powerful Gramian, \eg 10-order Gramian as used in the work~\cite{sastry2020detecting}, see details in Section~\ref{sec: Effectiveness Against Dynamic Backdoor}) in detecting trojaned inputs.

\noindent \textbf{Our work.}  
In this paper, we show that Gramian information of dynamically trojaned data points is highly distinct from that of the benign ones. 
Therefore, if we carefully design the order information (\eg less than 10-order) and detection metrics, a Gram matrix could be an effective tool for backdoor detection.

Our method, \textit{\textbf{Beatrix}} (\underline{b}ackdoor d\underline{e}tection via gr\underline{a}m ma\underline{trix}), captures not only the feature correlations but also the order information of the intermediate representations to reveal subtle changes in the activation pattern caused by backdoor triggers. 
Beatrix learns robust class-conditional statistics from the activation patterns of legitimate samples to effectively and efficiently harness the Gramian information in trojan detection. 
In the presence of a backdoor attack, Beatrix can capture the anomalies in the activation patterns since the difference in the feature representations of poisoned samples and legitimate samples is highlighted by our detection metrics.

\noindent \textbf{Contributions.} Our main contributions are summarized as follows.
\begin{itemize}[leftmargin=*]

\item We present a comprehensive analysis and insights of main-stream defenses to unveil their limitations against dynamic backdoor attacks.

\item We develop and implement Beatrix, a novel approach to defend against backdoor attacks. Beatrix utilizes a statistically robust deviation measurement with Gramian information to capture the anomalies in the activation patterns induced by poisoned samples. Beatrix also leverages Regularized Maximum Mean Discrepancy to further improve the performance in identifying infected classes. 
\item We demonstrate the effectiveness and robustness of our proposed method through theoretical justifications and extensive comparisons with state-of-the-art defensive techniques. We show that Beatrix can effectively detect sample-specific backdoor attacks and significantly outperform the existing defenses.

\end{itemize}

\section{Background}
\label{sec: Background}
In this section, we begin by briefly introducing the concept of Gram Matrix and the advances in backdoor attacks. We then discuss the limitations of existing defenses.

\subsection{Gram Matrix in DNNs}
Gramian information is widely used in areas such as Gaussian process regression~\cite{rasmussen2003gaussian} and style transfer learning~\cite{gatys2016image}. It computes the inner products of a set of $m$-dimensional vectors. The vectors, for instance, can be random variables in Gaussian process or vectorized internal activation patterns in style transfer. 
Formally, we suppose $A:=\{a_k|a_k\in\mathbb{R}^{m}\}_{k=1}^{n}$ is a set of $m$-dimensional random variables, and then the gramian information between $a_i, a_j\in A$ is defined as $G_{ij} = \sum_{k}^{m} a_{ik} \cdot a_{jk}$.
Thus, the Gram matrix $G$ is an $n \times n$ symmetric matrix containing the gramian information between each pair of random variables in $A$.
Since the off-diagonal entries of $G$ represent the pairwise correlation between $a_{i}$ and $a_{j}$, Gram matrix can be used as an covariance matrix in Gaussian process regression \cite{rasmussen2003gaussian}. On the other hand, due to its effectiveness in feature learning, the Gram matrix shows remarkable performance in capturing stylistic attributes (\textit{e.g.}, textures and patterns) in neural activations~\cite{li2017demystifying}. 
The high-order form of the Gram matrix has also been leveraged to improve OOD detectability~\cite{sastry2020detecting}.
The entries of $p$-th order of the Gram matrix is defined as $G_{ij}^{p} = \left(a_{i}^{p}{a_{j}^{p}}^{T}\right)^{1/p}$, where $p$ is the exponent.

\subsection{Backdoor Attacks}

Backdoor attacks are a technique of injecting some hidden malicious functionality into ML systems~\cite{pang2022trojanzoo, gao2020backdoor, liu2020survey}. The injected backdoor is activated only when a certain trigger appears in the input. This hidden functionality usually results in misclassifying the input sample into a target class predefined by the attacker. 

\noindent \textbf{Universal (sample-agnostic) backdoor.~}
Although various backdoor attacks~\cite{gu2019badnets, chen2017targeted, yao2019latent, jia2022badencoder} have been proposed, the majority of them have a static trigger setting, meaning that there is only one universal trigger and any clean sample with that trigger will be misclassified to the target label~\cite{li2020rethinking}.
Particularly, in the most common backdoor attack (\textit{i.e}., BadNets~\cite{gu2019badnets}), an adversary can form a backdoor trigger $t=(m,p)$, where $m$ and $p$ denote the blending mask and the trigger pattern, respectively. 
During the training of a DNN, a clean training sample pair such as $(x,y)$ is randomly replaced by the poisoned pair $(x_{bd}, y_{bd})$ with a certain probability using the trigger embedding function $\mathcal{B}$, which is defined as
\begin{align}
x_{bd} &= \mathcal{B}(x,t) \\
&= x\cdot(1-m) + p\cdot m
\end{align}

\noindent \textbf{Partial (source-specific) backdoor.~} 
In the partial attack, only samples in a specific source class can activate the backdoor and be misclassified into the target class set by the trigger~\cite{wang2019neural, tang2021demon}. As for samples in other classes, the trigger will not activate the backdoor. It is worth noting that all the trojaned samples still share the same uniform trigger in the source-specific backdoor attack. 

\noindent \textbf{Dynamic (sample-specific) backdoor.~}
Compared to the universal and partial backdoor attacks, dynamic backdoor attacks~\cite{salem2020dynamic, nguyen2020input, li2021invisible} make triggers that vary from sample to sample and this complicates the detection of such backdoors. 

To the best of our knowledge, there exist three dynamic backdoor attacks~\cite{salem2020dynamic, nguyen2020input, li2021invisible}~\footnote{We include~\cite{li2021invisible} here because it is a sample-specific backdoor attack which is very similar to the input-aware dynamic attack of \cite{nguyen2020input}, though the authors do not use the term \textit{dynamic} in their paper~\cite{li2021invisible} explicitly.}. All of them utilize a trigger generating network to launch dynamic backdoor attacks. 
In this work, in addition to universal backdoors, we try defending against the state-of-the-art dynamic ones, and more specifically, invisible sample-specific ~\cite{li2021invisible} and input-aware dynamic backdoor~\cite{nguyen2020input} attacks. (We do not consider~\cite{salem2020dynamic} since its triggers are not sample-specific and the code is not released.) 
Both of them consider the uniqueness and exclusiveness of triggers, \textit{i.e.}, each sample has a unique trigger which is non-reusable for any other sample. Herein, we will brief their attack paradigms, and readers can find more details from the original papers~\cite{nguyen2020input, li2021invisible}. 

Compared to the fixed and universal backdoor triggers, a \textit{dynamic} or \textit{sample-specific} trigger is a function of the corresponding input sample $x$. Suppose $g$ is a trigger generator. It can be defined as a function mapping an input from the sample space $\mathcal{X}$ to a trigger in the trigger space $\mathcal{T}$:
\begin{align}
g: & x\in\mathcal{X} \rightarrow t\in\mathcal{T}.
\end{align}
Additionally, the dynamic triggers should be non-reusable and unique. Henceforth,
\begin{equation}
    \mathop{\arg\max}_{y^{*}} f_{y^{*}}(\mathcal{B}(x,g(\hat{x}))) = y_{bd}\cdot\mathds{1}(x = \hat{x}) + y\cdot\mathds{1}(x \neq \hat{x}),
\end{equation}
where $y_{bd}$ is an backdoor target and $y$ is the ground truth label of $x$. In other words, a clean sample $x$ with the trigger generated based on another sample will not activate the hidden backdoor in the model $f$.

Both dynamic backdoor attacks and adversarial attacks aim to make models misbehave and share many similarities. Still, they have certain differences.
Given a classifier $f$, an adversary injects dynamic backdoor by jointly training a trigger generation function $g$ with $f$ on a clean distribution $p_{data}$ and a poisoning distribution $p_{bd}$. The overall training objective is 
\begin{equation}
    \max_{g, f}\Pr_{(x,y_{bd})\sim p_{bd}}[f(x+g(x))=y_{bd}] + \max_{f}\Pr_{(x,y)\sim p_{data}}[f(x)=y].
\end{equation}

The training objective may recall a similar objective in targeted evasive (adversarial) attacks in which an adversary $\hat{g}: \max_{\hat{g}}\Pr[f(x+\hat{g}(x))=y_t]$. However, it is easily found that the dynamic backdoor attacker has the capability of training $g$ on the training dataset of $f$ while the adversarial attacker optimizes $\hat{g}$ either in a per-sample manner or over an external dataset whose distribution is similar to that of the training dataset. As a ramification, $f_{y_{bd}}(x+g(x)) \gg f_{y^*:y^*\neq y_{bd}}(x+g(x))$, which means $f$ tends to be overfitted to the data distribution modified by $g(x)$. Such attack was found hard to be detected by adversarial detection methods due to the statistics of $f$ has been changed and the high confidence in the misclassification of $x+g(x)$~\cite{athalye2018obfuscated}. 
In addition, further modifying the triggers during the inference stage harms the stealthiness of the dynamic backdoor attacks and requires more capability from the attacker's side, which breaks the threat model of the dynamic backdoor. Such modification generates a separate adversarial attack rather than being a part of the dynamic backdoor attack, which falls out of the scope of this paper.

\subsection{Existing Defenses}
\label{sec: Limitations against Dynamic Backdoors}

\begin{table}[t]
\caption{\label{table:SoL}Limitations against Dynamic Backdoors}
\centering
\resizebox{1.\linewidth}{!}{%
\begin{tabular}{ll}
\toprule
\multicolumn{1}{c}{\textbf{Defense}} & \multicolumn{1}{c}{\textbf{Limitation}} \\
\midrule
Type-I  & \begin{tabular}[c]{@{}l@{}}Perturbation-resistant assumption of triggers\end{tabular}\\
Type-II & \begin{tabular}[c]{@{}l@{}}Can only reconstruct sample-agnostic triggers\end{tabular} \\
Type-III & \begin{tabular}[c]{@{}l@{}}Strong assumption on the distribution of feature representations\end{tabular}\\
\bottomrule
\end{tabular}
}
\end{table}

As summarized in Table~\ref{table:SoL}, Type-\uppercase\expandafter{\romannumeral1} defenses assume that the backdoor trigger is resistant to perturbations. Thus, the trigger regions or trigger-carrying images can cause the same misclassification when overlaid on other clean images~\cite{gao2019strip, chou2020sentinet, doan2020februus}. However, this assumption is violated in dynamic backdoors where the trigger is only activated for a specific sample. Moreover, Type-\uppercase\expandafter{\romannumeral2} defenses try to reconstruct a universal trigger that can convert any clean sample to the same target class~\cite{wang2019neural, liu2019abs}. Unfortunately, this reconstructed trigger is only valid for sample-agnostic backdoor. In contrast, the dynamic backdoor triggers vary from sample to sample, rendering the reconstructed universal trigger to be  totally different from the actual dynamic triggers. In addition, Type-\uppercase\expandafter{\romannumeral3} defenses attempt to distinguish the difference between the representations of clean samples and those of trojaned samples. However, they either use trivial clustering techniques~\cite{chen2019detecting, tran2018spectral} or model the representations with Gaussian distributions~\cite{tang2021demon, hayase2021spectre}, which cannot resist dynamic backdoor attacks. The detailed results can be found in our experimental analysis in Section~\ref{sec: comparison}.

\noindent \textbf{A case study of SCAn.} 
Although SCAn~\cite{tang2021demon} reveals the drawbacks of current defenses relying on the sample-agnostic backdoor assumption, \textit{it only considers the partial backdoor but leaves the sample-specific attack as an open problem}. We argue that there are three limitations of SCAn, which indicates SCAn cannot be extended to defending against the sample-specific backdoor.  
Firstly, SCAn assumes that the representations of normal and trojaned samples can be distinguished by the first moment (mean) discrepancy (\textit{Two-component decomposition assumption} in SCAn). However, in the dynamic backdoor attack, the first moment information becomes less discriminative. 
Secondly, SCAn models the feature distribution by a Gaussian distribution under the Linear Discriminant Analysis (LDA)~\cite{mika1999fisher} assumption, \textit{i.e.}, different mean values but same covariance for the distributions of clean and trojaned feature representations (\textit{Universal Variance assumption} in SCAn). 
However, our normality test shown in Figure~\ref{fig:normality_test} resonates with the observation of previous work~\cite{zisselman2020deep} that the feature space of DNNs does not necessarily conform with a Gaussian distribution. 
Thirdly, as demonstrated by our theoretical analysis in Appendix~\ref{appendix: Theoretical  Analysis of SCAn}, the essence of SCAn is to compute the weighted Mahalanobis distance between the representations of clean samples and those of trojaned samples, indicating that the effectiveness of SCAn is also dependent on the weighted value (the ratio of trojaned samples).
According to their estimate~\cite{tang2021demon}, SCAn needs to discern roughly 50 trojaned images before it can reliably detect further attacks. This is a severe drawback for security-critical applications. For example, the adversary may have bypassed an authentication system dozens of times before being caught.

\section{Overview and Motivation}
\label{sec: Overview}

To overcome the limitations of existing defenses, we propose a novel backdoor detection approach that covers the sample-specific backdoor attack. We first introduce the threat model in our work and then present our key observations and ideas.
Finally, we analyze the superiority of Gram matrix.

\subsection{Threat Model}

We consider the standard threat model which is consistent with that of the most recent backdoor attack and defense studies~\cite{gu2019badnets, wang2019neural, tang2021demon}.

\noindent \textbf{Adversary.~}
Similar to most backdoor poisoning settings, the goal of the adversary is to deliberately inject one or more backdoors into the target model. 
The compromised model performs well on clean samples, whereas it misclassifies attack samples (trigger-carrying samples) to the predefined target label. We assume that the adversary can access the training set of the model, and is capable of poisoning the training data without major constraints, but has no direct access to the model. This scenario allows us to study the attack under worst-case conditions from the defender's point of view.

\noindent \textbf{Defender.~} The goal of the defender is to perform input-level detection to determine whether an input will trigger a malicious behavior in a untrusted model in an online setting such as the Machine-Learning-as-a-Service scenario. Furthermore, the defender aims to tell the infected classes of a backdoored model based on the instances it classifies. 
We assume that the defender has white-box access to the target model, including the feature representation in the intermediate layers. Additionally, the defender needs a small set of clean data to help it with the detection, which was also a requirement in the previous works~\cite{chou2020sentinet,gao2019strip,tang2021demon, xu2021detecting}.

\subsection{Intuition and Key Idea}
\label{sec: Intuition and Key Idea}

\begin{figure}[t]
\centering
\includegraphics[width=0.3\textwidth]{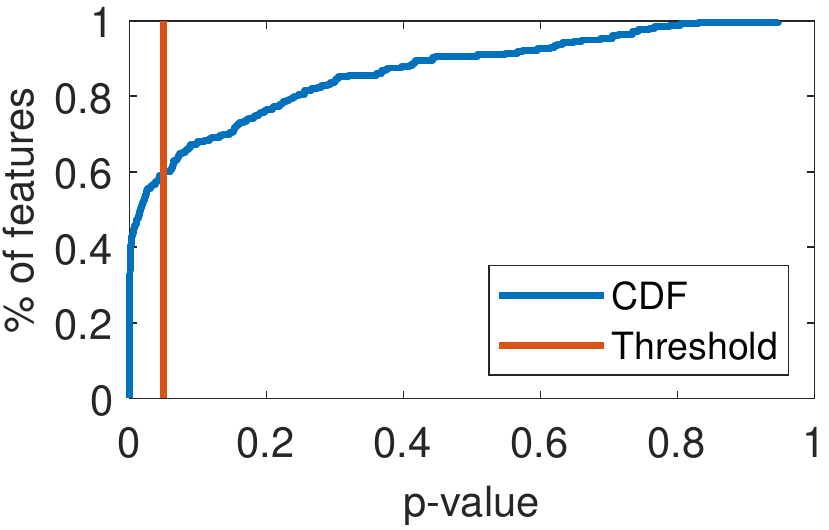}
\caption{\label{fig:normality_test} Normality Test by Shapiro-Wilk test. We can find that about 60\% features do \textbf{NOT} follow a normal distribution under a 95\% confidence score. This demonstrates that the Gaussian distribution assumption in \cite{tang2021demon,hayase2021spectre} is untenable in more advanced attacks such as the dynamic backdoor attack.}
\end{figure}

\begin{figure*}[t]
\centering
\includegraphics[width=0.88\textwidth]{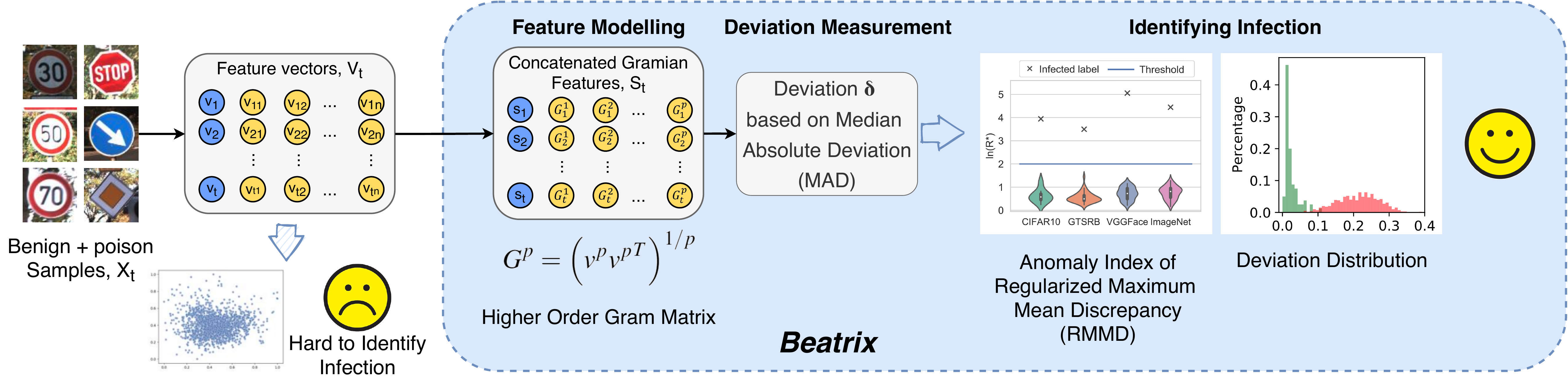}
\caption{\label{fig: workflow} An overview of Beatrix.}
\vspace{-7mm}
\end{figure*}

\noindent \textbf{Intuition.~} A key observation is that the clean samples of a certain class and the trojaned (trigger-carrying) samples targeted at that class are disjoint in the pixel space. 
Consequently, even though a trojaned sample is misclassified into the target label, its intermediate representation is somehow different from those of normal samples of the target class.
The anomaly triggered by the trojaned samples can be characterized by inconsistencies between the intermediate feature representations and their predicted labels. This observation provides a basis for investigating the problem of characterizing trojaned samples from the perspective of OOD detection.

A predictive uncertainty study~\cite{amodei2016concrete} reveals that DNNs perform well on the samples drawn from the distribution seen in the training phase, but tend to behave unexpectedly when they encounter OOD samples that lie far from the training distribution. 
Analogously, trojaned samples can be thought of as OOD samples drawn from a distinct distribution in contrast to the distribution of clean samples. Therefore, we believe there is a link between the detection of trojaned samples and the OOD detection~\cite{liang2017enhancing, lee2018simple, zisselman2020deep, sastry2020detecting}.

However, we notice there exist differences between OOD detection and backdoor detection. 
First, the feature representations of OOD samples can be effectively modeled by Gaussian distributions. This assumption shows superior performance on OOD detection tasks~\cite{lee2018simple, zisselman2020deep}. On the contrary, Gaussian distributions are less capable of modeling the features of trojaned samples, due to the complexity and diversity of backdoor triggers. This dilemma is further emphasized by dynamic backdoors. As our experimental results show in Section~\ref{sec: comparison}, backdoor detection methods that use Gaussian distribution, \textit{e.g.}, SCAn~\cite{tang2021demon}, achieve suboptimal performance in the identification of dynamic backdoors. A normality test in Figure~\ref{fig:normality_test} also exposes the problem. 
Second, less attention has been paid to adversarial robustness in OOD detection methods. \textit{A priori} knowledge of in-distribution/clean samples is a key piece of information for many OOD detectors~\cite{lee2018simple, zisselman2020deep, sastry2020detecting} and backdoor detection methods~\cite{gao2019strip, wang2019neural}. The OOD detection task assumes that a set of clean samples used for training the detector can be well curated. However, this assumption is challenged in the scenario of backdoor detection since poisoned samples may carry invisible triggers~\cite{li2021invisible, shafahi2018poison} which can hardly be filtered even by manual inspection. 
The lack of adversarial robustness restricts the deployment of OOD detectors in detecting trojaned inputs (see~\ref{sec: Effectiveness Against Dynamic Backdoor} in more detail). 
Finally, OOD detectors require sufficient in-distribution data, or even OOD data~\cite{liang2017enhancing, lee2018simple, zisselman2020deep}. This requirement ensures that the parameters of the detection models can be effectively estimated. However, there are usually limited clean data available to the backdoor defender. Thus, the requirements for statistical robustness under limited data are different in OOD and backdoor detection methods.

\noindent \textbf{The key idea.~}
From our observation, backdoor defense can be viewed as the problem of detecting OOD. We argue that the problem of finding a robust detector for neural backdoors can be connected to feature modeling methods used in areas such as Gaussian process regression, style transfer, and OOD detection~\cite{rasmussen2003gaussian, gatys2016image, sastry2020detecting}. In this paper,
we find that although the trojaned and clean samples are deeply fused in the original feature space, they are distinguishable in the Gramian feature space, indicating Gram matrix is an effective tool for feature modeling. The Gram matrix derives from the inner products of feature maps across different channels.
Thus, Gram matrices not only consider features in each individual channel but also incorporate the feature correlations across channels~\cite{sastry2020detecting,li2017demystifying}.

We note that the previous representation-based backdoor detection methods either use trivial clustering techniques~\cite{chen2019detecting, tran2018spectral} or Gaussian models~\cite{tang2021demon, hayase2021spectre} in the detection. 
These methods ignore the high-order information and only consider the first moment (mean) discrepancy between clean and trojaned samples. However, the simplification reduces the discriminative power of the methods against more complex attacks such as dynamic backdoors. 
To tackle the problem, we turn to high-order statistics of the feature representations since they are more sensitive to the changes in the feature space, due to its higher-power format. Our method employs not only first-order moments but also high-order moments for feature modeling. 
We utilize the Gram matrix and its appropriately high-order forms to capture not only the feature correlations but also appropriately high-order information to detect trojaned samples. 
In addition, considering the adversarial robustness and statistical robustness in backdoor detection, we do not use the multivariate Gaussian to model the trojaned samples in the deviation measurement as Gaussian models only perform well with sufficient data~\cite{psutka2015sample}. Instead, we utilize Median Absolute Deviation (MAD), a more robust estimation of statistical dispersion, to measure the deviation of trojaned samples. 
Finally, when the training dataset of a given class is contaminated, the set of examples can be viewed  as a mixture of two subgroups~\cite{chen2019detecting, tran2018spectral, tang2021demon, hayase2021spectre}. In contrast to previous works~\cite{tang2021demon, hayase2021spectre} assuming that the two subgroups follow Gaussian distributions with two different means but the same covariance, we employ Regularized Maximum Mean Discrepancy (RMMD)~\cite{danafar2013testing} to enhance the adversarial robustness of our method. RMMD is a Kernel-based two-sample testing technique which does  not have any assumption on the distributions. RMMD performs a hypothesis test on whether the feature representations in a given class are drawn from a mixture group (\textit{i.e.}, contaminated class) or a single group (\textit{i.e.}, uncontaminated class).

\subsection{Theoretical  Analysis of Gram Matrix}\label{appendix: Theoretical  Analysis of Gram Matrix}
Let $x^p\sim P_{c,p}$ and $y^p\sim P_{t,p}$ be $p$-th order representation vectors sampled from the clean distribution $P_{c,p}$ and the trojaned distribution $P_{t,p}$, respectively. Beatrix models the feature representations without any assumption on $P_{c,p}$ and $P_{t,p}$. Instead, Beatrix relies on Gram matrix to extract discriminative information of $P_{c,p}$ and $P_{t,p}$ from their statistical moments~\cite{bishop2006pattern}. 
Let $u^p_1$ and $S^p_1$ be the mean vector and the covariance matrix of $x^p$, respectively. The second raw moment of $x^p$ over $P_{c,p}$ is:
\begin{align}
E({x^p}{x^p}^T) &= E({x^p})E({x^p})^T + E[({x^p}-u^p_1)({x^p}-u^p_1)^T] \\\nonumber
&= u^{p}_{1}{u^p_{1}}^T + S^p_1.
\end{align}
The Gramian feature of ${x^p}$ is defined as $G_{x^p} = {x^p}{x^p}^T$. Then, $E(G_{x^p}) = E({x^p}{x^p}^T) = u^p_{1}{u^p_{1}}^T + S^p_1$.
Similarly, let $u^p_2$ and $S^p_2$ be the mean vector and the covariance of $y^p$, we have $E(G_{y^p}) = u^p_{2}{u^p_{2}}^T + S^p_2$
over $y^p\sim P_t$.
Therefore, the expected discriminative information captured by the Gram matrices over different exponents can be represented from the prospective of statistical moments:
\begin{align}
\label{equ: moment discrimination captured by Gram matrix}
M(P_c,P_t) &= E_{p\in\mathbb{Z^+}} [E(G_{x^p}) - E(G_{y^p})]\\\nonumber
&= E_{p\in\mathbb{Z^+}} [u^{p}_{1}{u^p_{1}}^T - u^{p}_{2}{u^p_{2}}^T + S^p_1 - S^p_2],
\end{align}
where $P_c$ is a collection of distributions of clean representations with elements of different powers and $P_t$ is that of trojaned representations.
Equation~\ref{equ: moment discrimination captured by Gram matrix}
shows that Gramian features not only capture the first moment discrepancy (\textit{i.e.}, $u^p_{1}{u^p_{1}}^T - u^p_{2}{u^p_{2}}^T$) like SCAn (when $p=1$, see Equation~\ref{equ: scan likelihood ratio}), but also the second moment discrepancy (\textit{i.e.}, $S^p_1 - S^p_2$).
Compared to previous methods such as SCAn, a trojaned $y$ can still be distinguished from clean representations when the clean and trojaned distributions have the same mean. Moreover, by considering various $p$ values, the second moment discrepancy can capture more information about high-order features and better models $M(P_c,P_t)$.

\section{Design of Beatrix}
\label{sec: Design}

In this section, we provide the details of our approach to detecting backdoor attacks. The framework of Beatrix is illustrated in Figure~\ref{fig: workflow}.

\subsection{Feature Modeling via Gram Matrices}\label{subsec:gramian}

Formally, let $h$ be the sub-model up to the $l$-th layer of a DNN model. Then, the feature representation of the input sample $x$ at the $l$-th layer of the DNN model is defined as $h(x) = v \in \mathbb{R}^{n \times m}$, where $n$ is the number of channels at layer $l$ and $m$ is the height times the width of each feature map. The features correlations between channels can be expressed by
\begin{align}
G = vv^{T}, 
\end{align}
where $G \in \mathbb{R}^{n\times n}$ denotes the Gram matrix of the feature maps in an inner product space. 
In order to capture more prominent activations in feature maps, we also use the high-order Gram matrix
\begin{align}
G^{p} = \left(v^{p}{v^{p}}^{T}\right)^{1/p}, 
\end{align}
where $v^{p}$ denotes the $p$-th power of the feature representation~$v$, and $G^{p}$ is the $p$-th order Gram matrix of $v$.

The off-diagonal entries of $G^{p}$ represent the pairwise correlation between feature maps at the $l$-th layer while the entries at the diagonal only relate to a single feature map. 
Since the matrix $G^{p}$ is symmetric, we only need the upper (or lower) triangular part of it. In particular, the vectorized triangular matrix which contains the entries on and above (or below) the main diagonal, can form a $\frac{1}{2}n(n+1)$-dimensional vector like $\Vec{G^{p}}$.

We can compute $\Vec{G^{p}}$ for each order $p \in \{1,...,P\}$, where $P$ is a hyperparameter representing the bound of the order. 
By concatenating all the output vectors $\Vec{G^{p}}$, we can derive a new representation vector $s = [\Vec{G^{1}},\  \Vec{G^{2}}, \ ..., \ \Vec{G^{P}}] \in \mathbb{R}^{\frac{1}{2}n(n+1)P}$ for the input sample $x$.

Let $\mathcal{X}_t$ denote a set of clean samples in class $t$. $\mathcal{X}_t$ has a feature presentation set $\mathcal{V}_t = \left\{ v_i:= h(x_i),\ x_{i} \in \mathcal{X}_t \right\}$ and a concatenated Gramian feature set $\mathcal{S}_t = \left\{s_i,\ i \in \{1,2,...,|\mathcal{X}_t|\} \right\}$. The task of backdoor detection can be formulated as an outlier detection problem: given the feature $\hat{s}$ of an input sample $\hat{x}$ and its predicted label $\hat{y_t}$ by the target model $f$, we try to determine whether $\hat{s}$ is an outlier, with respect to $\mathcal{S}_t$ based on the statistical properties of $\mathcal{S}_t$. 

\subsection{Deviation Measurement}\label{subsec:dm}

A natural choice of computing the deviation of a point $\hat{s}$ is to build a multivariate Gaussian model of $\mathcal{S}_t$. However, the problem is non-trivial because of \textit{i)} the large dimensionality of the feature vector $s$ and \textit{ii)} the limited number of clean samples for estimating Gaussian parameters (especially, the covariance matrix). 
Therefore, building a multivariate Gaussian model for high dimensional variables is not statistically robust when there are limited data samples available. 
Additionally, since the feature modeling with Gram matrices has already considered the feature correlations, we can just simplify the problem and model each element in $\mathcal{S}_{t}$ independently. Thus, high dimensional estimation can be simplified into one dimensional estimation for each independent element. 

In the simplified case, one may still consider using the Gaussian model but its univariate version to estimate mean and standard deviation of the features. However, recall that there is limited clean data available for the defender, the estimation results (\textit{i.e.}, mean and standard deviation) are easier to be affected by the outliers as Gaussian models only perform well with sufficient data~\cite{psutka2015sample}. More importantly, the individual elements of $s$ may not follow a Gaussian distribution strictly.

Instead of using a univariate Gaussian model, we propose to utilize Median Absolute Deviation (MAD) which is known to be more resilient to outliers in a dataset than the standard deviation $\hat{\sigma}$~\cite{leys2013detecting}. The absolute deviations between all data points and their medians are gained before MAD is employed as the median of these absolute deviations. 

Given the set of concatenated Gramian features $\mathcal{S}_t$ of all clean samples $\mathcal{X}_t$ in class $t$, we can compute the median and the MAD with respect to each concatenated Gramian vector:
\begin{align}
\tilde{s}_j =& \  median(\{s_{ij},\ \forall i\in\{1,2,...,|\mathcal{X}_t|\}\}), \\
MAD_{j} =&\! \ median(\{|s_{ij}\!-\!\tilde{s}_j|,\!\ \forall i\!\in\!\{1,2,...,|\mathcal{X}_t|\}\}).
\end{align}
Then the deviation of the observed $j$-th value $\hat{s}_j$ in the candidate feature point $\hat{s}$ is defined as:
\begin{align}
\delta_j &= \delta(\hat{s}_j) \\
&=\left\{
\begin{aligned}
0\quad\quad\quad,&\quad if \quad min \leq \hat{s}_j \leq max, \\
\frac{min-\hat{s}_j}{min},&\quad if \quad \hat{s}_j \leq min, \\
\frac{\hat{s}_j-max}{max},&\quad if \quad max \leq \hat{s}_j, 
\end{aligned}
\right.
\end{align}
where $min=\tilde{s}_j - k\cdot MAD$, $max=\tilde{s}_j + k\cdot MAD$, and $k$ is a predefined scale factor that is set to 10 in our case.

Then the deviation of the candidate feature point $\hat{s}$ is the sum of the deviation values over all entries in $\hat{s}$:
\begin{align}
\delta = \frac{2}{n(n+1)P}\sum_{j=1}^{\frac{1}{2}n(n+1)P} \delta_j.
\end{align}

\vspace{-1mm}
\noindent \textbf{Threshold determination.~}
The detection boundary of Beatrix is estimated by benign inputs. Due to the limited number of clean data available to the defender, we employ bootstrapping to compute the deviations of benign inputs~\cite{efron1994introduction}. Specifically, we randomly draw $\frac{1}{T}$ samples from the clean dataset as testing samples. The remaining samples are used as training samples to estimate the min/max values. The procedure is repeated for $T$ iterations to obtain the deviations of benign samples. 
The detection boundary can be determined by the defender when choosing different percentiles like STRIP~\cite{gao2019strip}. 
For example, the defender can choose 95\% as the detection boundary. This means that 95\% of the deviations of benign samples are less than this detection boundary.

\subsection{Identifying Infected Labels}
The performance of Beatrix can be further improved through a local refinement of the detection results to reduce false positives. Since the detection threshold in Section~\ref{subsec:dm} is predefined, there could be false positive in the detection results. In an offline setting, Beatrix accumulates the historical detection results for a second statistical analysis to ablate false trojaned targets given by the threshold thereof.
In the presence of a backdoor attack, the feature representations of samples in the infected class can be considered as a mixture of two subgroups~\cite{chen2019detecting, tran2018spectral, tang2021demon, hayase2021spectre}. However, previous works~\cite{tang2021demon, hayase2021spectre} assume that these two subgroups follow Gaussian distributions with two different means but the same covariance. Therefore, they perform a hypothesis testing to determine whether these two distributions are significantly different. However, as we discussed in Section~\ref{sec: Limitations against Dynamic Backdoors}, the Gaussian assumption is not tenable in more complex scenarios, such as dynamic backdoor attacks. 

Therefore, we resort to the Kernel-based two-sample testing which addresses whether two sets of samples are identically distributed without assumption on their distributions~\cite{gretton2012kernel, danafar2013testing, chwialkowski2015fast,  varoquaux2019comparing}.
A popular test statistic for this problem is the Maximum Mean Discrepancy (MMD)~\cite{gretton2012kernel}, which is defined based on a positive definite kernel function $k$~\cite{scholkopf2002learning}. Kernel methods provide the embedding of a distribution in a reproducing kernel Hilbert space (RKHS). For MMD with a linear kernel, $k(x, x^{\prime}) = \langle x, x^{\prime}\rangle$, it measures the distribution distance under their first moment discrepancy~\cite{muandet2016kernel}. In practice, a common option is to use the Gaussian kernel $k(x, x^{\prime}) = exp(-\beta\Vert x - x^{\prime} \Vert ^{2}_{2})$, which contains infinite order of moments by looking at its Taylor series~\cite{li2015generative}. 

In this work, we use an extension of MMD metric, termed Regularized MMD (RMMD), which incorporates two penalty terms to achieve better performance when the two sample sets are small and imbalanced~\cite{danafar2013testing}: 
\vspace{-1mm}
\begin{small}
\begin{align}
RMMD(P,Q) \!&=\! MMD(P, Q)^{2} \!-\! \lambda_{P} \Vert\mu_{P}\Vert_{\mathcal{H}}^{2} \!-\! \lambda_{Q} \Vert\mu_{Q}\Vert_{\mathcal{H}}^{2} \\
& = 
\! \Vert\mu_{P} - \mu_{Q}\Vert_{\mathcal{H}}^{2} \!-\! \lambda_{P} \Vert\mu_{P}\Vert_{\mathcal{H}}^{2} \!-\! \lambda_{Q} \Vert\mu_{Q}\Vert_{\mathcal{H}}^{2}, 
\end{align}
\end{small}
where $P$ and $Q$ denote the representation distributions represented by the samples in the two subgroups obtained from the deviation measurement.

According to the work~\cite{danafar2013testing}, this test statistic follows an asymptotic normal distribution based on theorems in~\cite{hoeffiding1948class, serfling2009approximation}. Similar to Neural Cleanse~\cite{wang2019neural} and SCAn~\cite{tang2021demon}, we can leverage the Median Absolute Deviation to identify the infected label(s) with abnormally large values of RMMD statistic instead of directly computing the p-value of this test. 
Specifically, we denote $R_{t}$ as the RMMD statistic of class $t$, and then anomaly index $R_{t}^{*}$ is defined as:
\vspace{-1mm}
\begin{align}
R_{t}^{*} &= |R_{t} - \tilde{R}|/(MAD(\tilde{R})*\eta),  \\
\text{where} \quad \tilde{R} &= median(\{R_{t}: t \in \mathcal{L}\}),\\
MAD(\tilde{R}) &= median(\{R_{t} - \tilde{R}|: t \in \mathcal{L}\}).
\end{align}

As $R_{t}$ follows an asymptotic normal distribution, we apply a constant factor $\eta = 1.4826$ to the anomaly index. We identify any label with an anomaly index $R_{t}^{*} \geq e^{2}$ as an infected label with the confidence probability $\geq (1 - 10^{-9})$~\cite{tang2021demon}.

\begin{table*}[t] 
\centering
\caption{\label{tabel: information of benigh models} Detailed information about dataset, model architecture and clean accuracy.}
\resizebox{0.98\textwidth}{!}{
\begin{tabular}{ccccccc}
\toprule
\textbf{Dataset}  & \# \textbf{of Classes} & \# \textbf{of Training Images} & \# \textbf{of Testing Images} & \textbf{Input size}    & \textbf{Model Architecture}  & \textbf{Top-1 accuracy} \\ \midrule
CIFAR10  & 10   & 50000 & 10000 & 32 $\times$ 32 $\times$ 3   & PreActResNet18 & 94.5\%  \\ 
GTSRB    & 43   & 39209 & 12630 & 32 $\times$ 32 $\times$ 3   & PreActResNet18 & 99.1\%  \\ 
VGGFace  & 100  & 38644 & 9661  & 224 $\times$ 224 $\times$ 3 & VGG16          & 90.1\%  \\ 
ImageNet & 100  & 50000 & 10000 & 224 $\times$ 224 $\times$ 3 & ResNet101      & 83.8\%  \\ 
\bottomrule
\end{tabular}}
\vspace{-3mm}
\end{table*}

\section{Evaluation}
\label{sec: Evaluation}
In this section, we first evaluate the effectiveness of our proposed method against the dynamic backdoor and then compare Beatrix with state-of-the-art defensive techniques. Finally, we also demonstrate the robustness of Beatrix against other attacks. The datasets and model structures used in our experiments are summarized in Table~\ref{tabel: information of benigh models}. We provide detailed introduction of the experiment setup in Appendix~\ref{sec: Experiment Setup}.

\subsection{Effectiveness Against Dynamic Backdoor}
\label{sec: Effectiveness Against Dynamic Backdoor}

\begin{figure}[t]
\centering
\subfigure[]{
\begin{minipage}[t]{0.065\textwidth}
\centering
\includegraphics[width=0.9\textwidth]{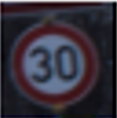}
\end{minipage}}
\subfigure[]{
\begin{minipage}[t]{0.065\textwidth}
\centering
\includegraphics[width=0.9\textwidth]{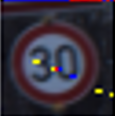}
\end{minipage}}
\subfigure[]{
\begin{minipage}[t]{0.065\textwidth}
\centering
\includegraphics[width=0.9\textwidth]{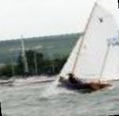}
\end{minipage}}
\subfigure[]{
\begin{minipage}[t]{0.065\textwidth}
\centering
\includegraphics[width=0.9\textwidth]{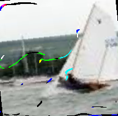}
\end{minipage}}
\subfigure[]{
\begin{minipage}[t]{0.065\textwidth}
\centering
\includegraphics[width=0.9\textwidth]{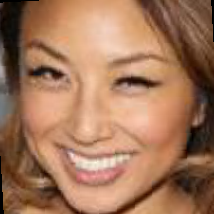}
\end{minipage}}
\subfigure[]{
\begin{minipage}[t]{0.065\textwidth}
\centering
\includegraphics[width=0.9\textwidth]{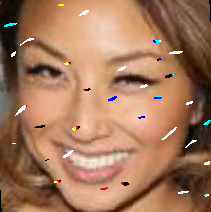}
\end{minipage}}
\caption{\label{fig: examples of dynamic} Examples of poisoned samples under the input-aware dynamic backdoor. (a), (c) and (e) are clean images, (b), (d) and (f) are poisoned images.}
\end{figure}

\noindent \textbf{Attack configuration.~} We implement the input-aware dynamic backdoor attack~\cite{nguyen2020input} using the code released by the authors~\cite{codeinputaware}. Figure \ref{fig: examples of dynamic} illustrates several examples of poisoned samples.
We conduct the common single-target attack, \textit{i.e.}, the target label is the same for all trojaned samples. We set both the backdoor probability (trojaned samples with their paired triggers) and the cross-trigger probability (trojaned samples with inconsistent triggers) as 0.1.
For all the four datasets, the backdoor attack success rates (ASR) are almost 100\% while still achieving a comparable performance on clean data as the benign models do, as shown in Table~\ref{tabel: accuracy of infected models}. It is worth noting that the cross-trigger accuracy (the accuracy of classifying images containing dynamic triggers deliberately generated for other images)
is over 80\% on all the four datasets, and this shows the nonreusability and uniqueness of the triggers on mismatched clean images. The evaluation results on the invisible sample-specific backdoor attack~\cite{li2021invisible} are shown in Section~\ref{sec: Robustness Against Other Attacks}.

\begin{table}[t]
\centering
\caption{\label{tabel: accuracy of infected models} Attack success rate, cross-trigger accuracy and classification accuracy of infected models.}
\resizebox{\linewidth}{!}{
\begin{tabular}{ccccc}
\toprule
\multirow{2}{*}{\textbf{Dataset}}& \multicolumn{3}{c}{\textbf{Infected Model}}  & \textbf{Benign Model}  \\
\cmidrule(r){2-4}\cmidrule(r){5-5}
& \begin{tabular}[c]{@{}c@{}}\textbf{Attack Success}\\ \textbf{Rate}\end{tabular} & \begin{tabular}[c]{@{}c@{}}\textbf{Cross-trigger}\\ \textbf{Accuracy}\end{tabular} & \begin{tabular}[c]{@{}c@{}}\textbf{Clean}\\ \textbf{Accuracy}\end{tabular} & \begin{tabular}[c]{@{}c@{}}\textbf{Clean}\\ \textbf{Accuracy}\end{tabular} \\ 
\midrule
CIFAR10  & 99.4\%    & 88.6\%  & 93.9\%   & 94.5\%  \\ 
GTSRB    & 99.7\%    & 96.1\%  & 99.2\%   & 99.1\%  \\ 
VGGFace  & 98.5\%    & 82.5\%  & 89.8\%   & 90.1\%  \\ 
ImageNet & 99.5\%    & 81.3\%  & 83.5\%   & 83.8\%  \\ 
\bottomrule
\end{tabular}}
\end{table}

\begin{figure*}[t]
\centering
\begin{minipage}[t]{0.24\linewidth}
\centering
\includegraphics[width=0.98\linewidth]{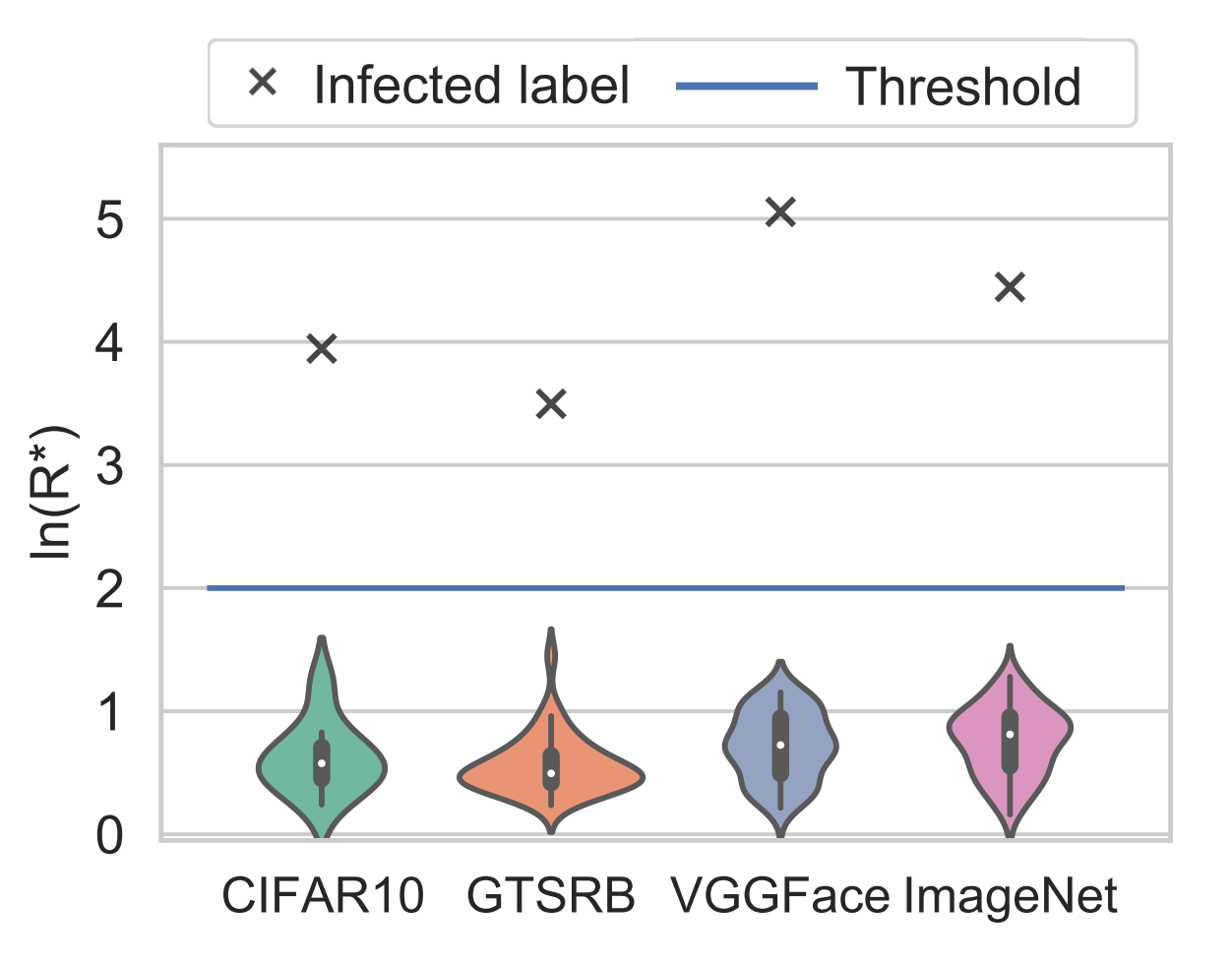}
\caption{\label{fig: detection results anomaly} The logarithmic anomaly index of infected labels on the four datasets.}
\end{minipage}
\quad
\begin{minipage}[t]{0.72\linewidth}
\centering
\includegraphics[scale=0.59]{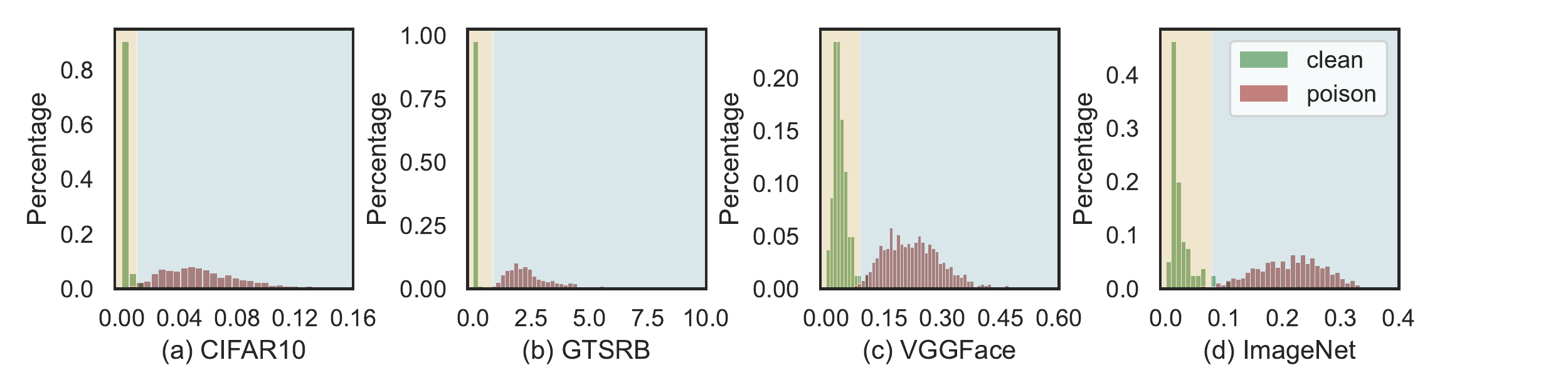}
\caption{\label{fig: detection results} Deviation distribution of benign and trojaned samples. The trojaned sample shows a much larger deviation than benign samples. The color boundary in the background indicates the decision threshold (same for the figures in the following sections).}
\end{minipage}
\vspace{-6mm}
\end{figure*}

\begin{figure*}[t]
\centering
\begin{minipage}[t]{0.3\textwidth}
\centering
\includegraphics[scale=0.43]{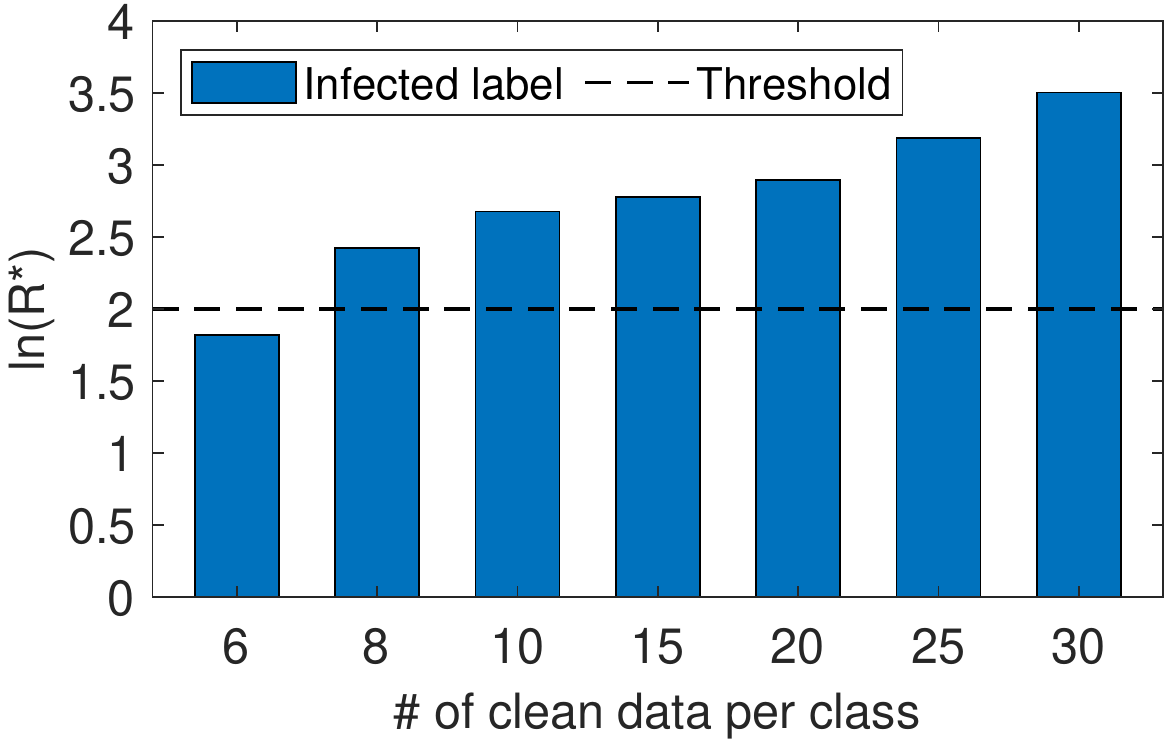}
\caption{\label{fig: clean_data_for_deviation} The logarithmic anomaly index of infected labels when using different numbers of clean data.
}
\end{minipage}
\quad
\begin{minipage}[t]{0.3\textwidth}
\centering
\includegraphics[scale=0.43]{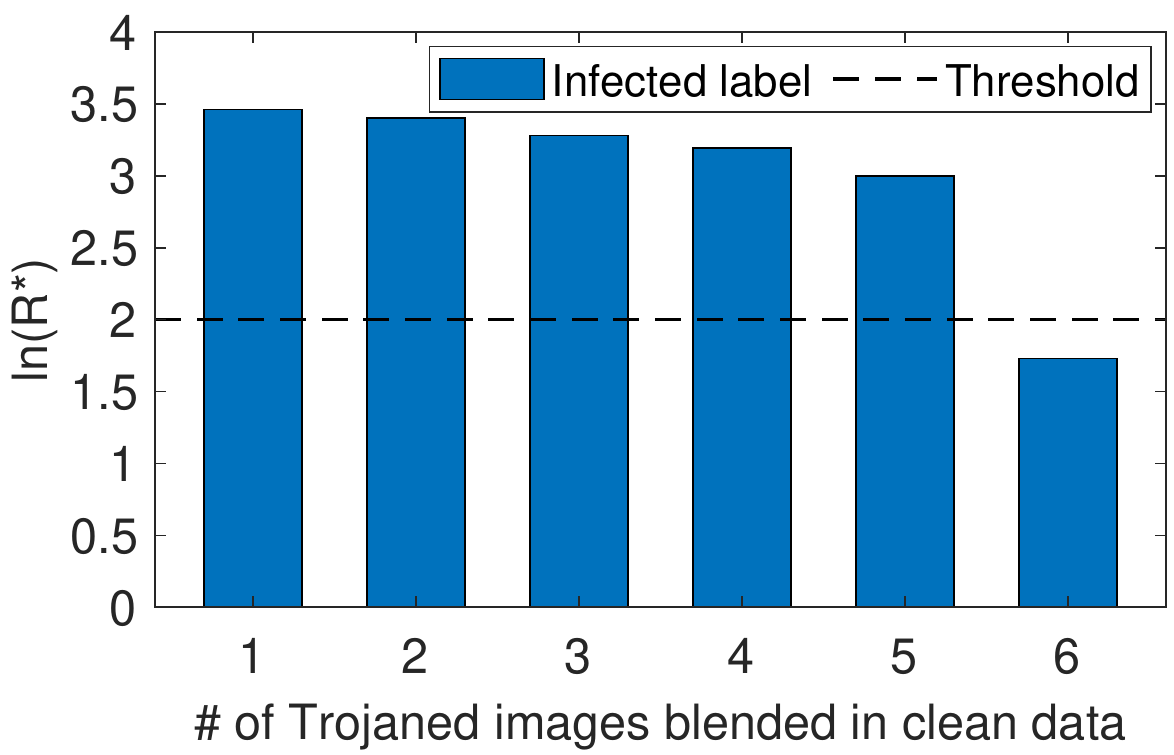}
\caption{\label{fig: clean_data_for_deviation_under_contamination} The logarithmic anomaly index of infected labels when clean data is contaminated.}
\end{minipage}
\quad
\begin{minipage}[t]{0.3\textwidth}
\centering
\includegraphics[scale=0.43]{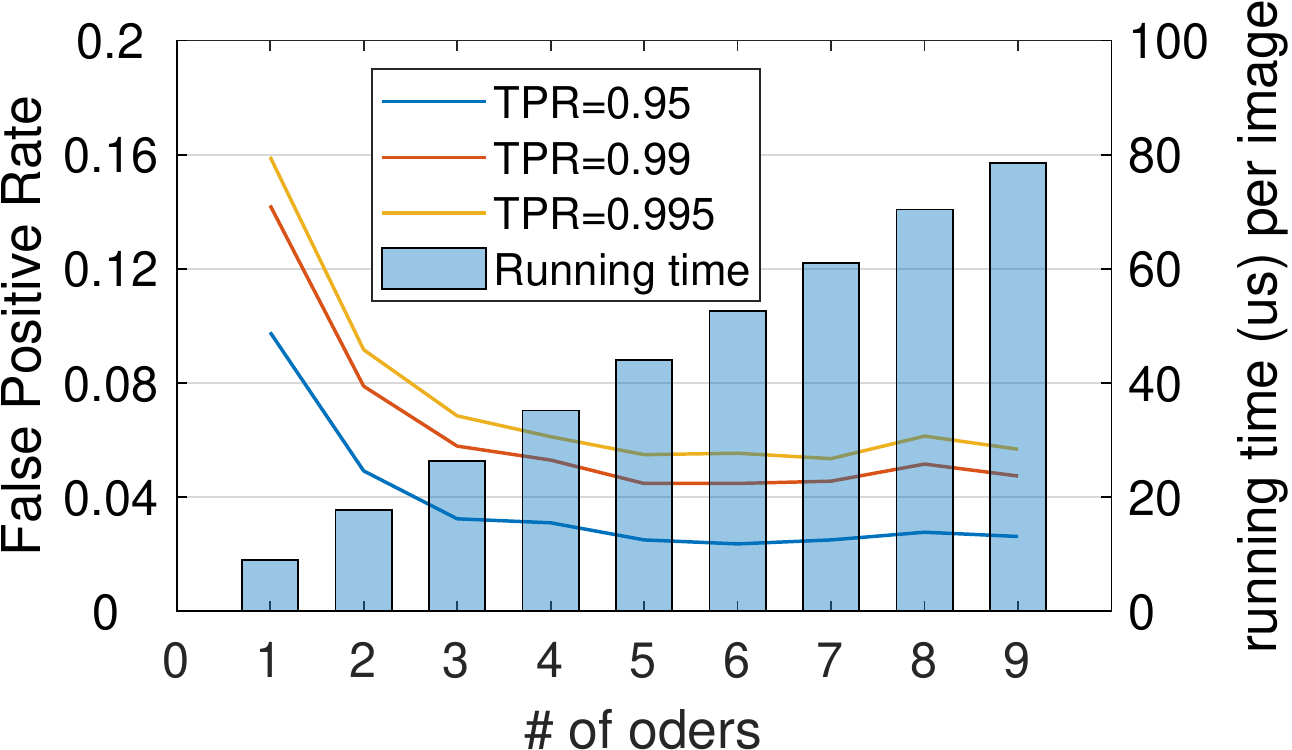}
\caption{\label{fig: order_for_deviation} False positive rate of benign images when incorporating different bound on the order of Gram matrix. 
}
\end{minipage}
\vspace{-7mm}
\end{figure*}

\noindent \textbf{Effectiveness on various datasets.~} From each dataset, we randomly select 30 images per class as a clean dataset for the defender. The clean dataset accounts for no more than 6\% of the whole dataset. 
The bound of the order of Gram matrix is set as 9. As is shown later, the detection effectiveness is stable when $P \geq 4$. 
The experimental results on the four datasets show that our defensive technique is very effective in detecting the dynamic backdoor attack. Figure~\ref{fig: detection results anomaly} illustrates the logarithmic anomaly index values $ln(R^{*})$ of infected and uninfected labels. All infected labels have much larger anomaly index values, compared to uninfected labels. This demonstrates that our defensive technique can effectively detect target classes in infected models on various datasets and model architectures. Moreover, Figure~\ref{fig: detection results} illustrates that the deviations of the trojaned samples are larger than those of the benign ones. This demonstrates that our method can also effectively distinguish benign from polluted samples.

\begin{figure*}[t]
\centering
\begin{minipage}[t]{0.26\linewidth}
\centering
\includegraphics[width=0.95\linewidth]{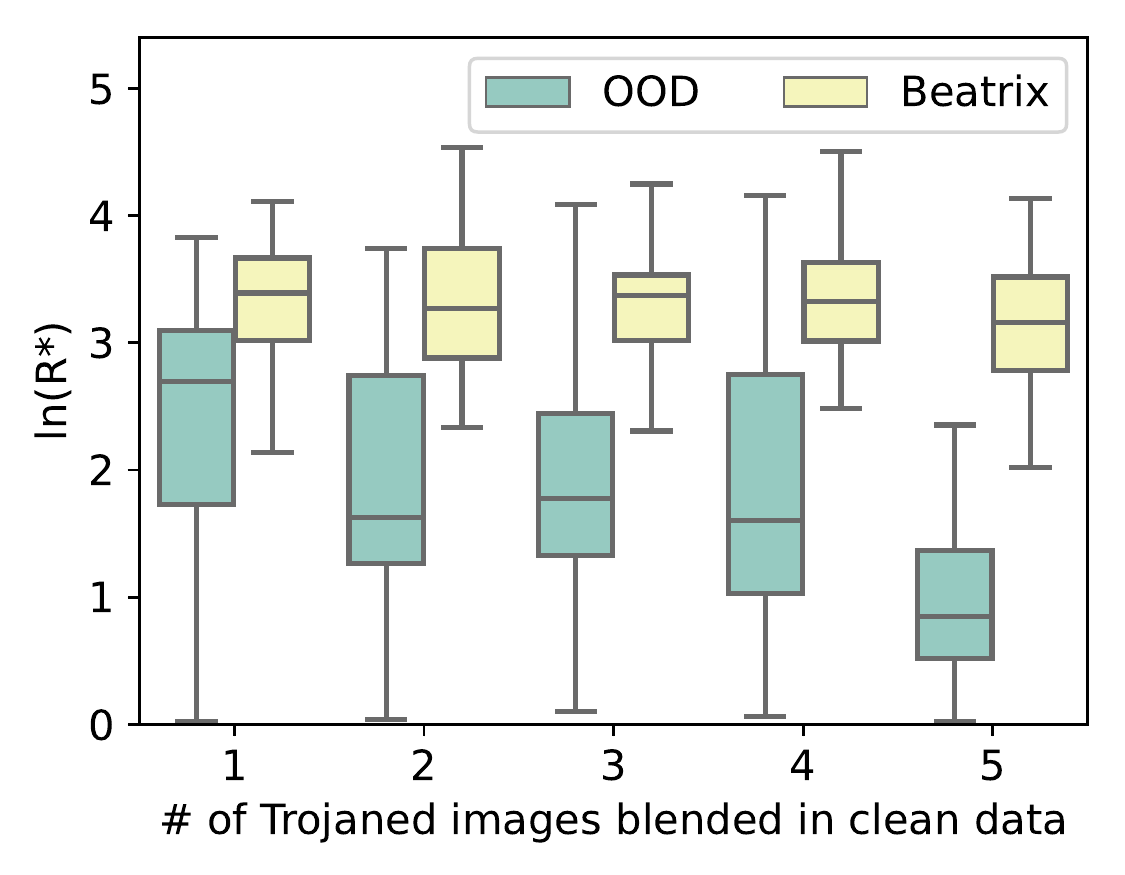}
\caption{\label{fig: ood vs beatrix} Robustness comparison between Beatrix and the OOD detection method \cite{sastry2020detecting}.}
\end{minipage}
\quad
\begin{minipage}[t]{0.68\linewidth}
\centering
\includegraphics[width=1\textwidth]{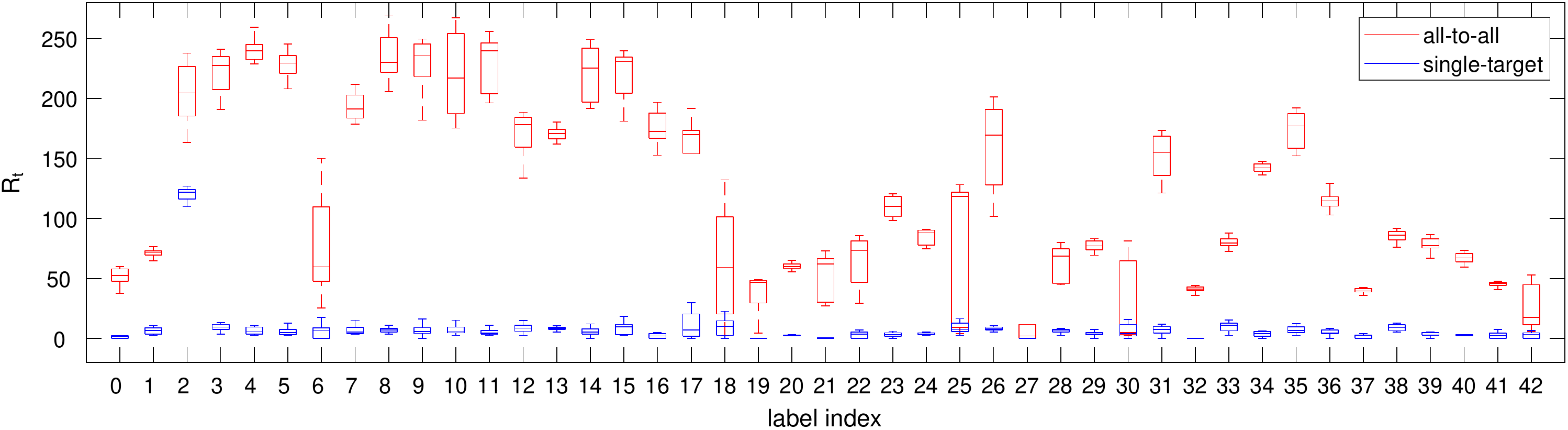}
\caption{\label{fig: all2all} RMMD statistics $R_{t}$ of different labels in all-to-all infected  models (red) and in  single-target (label 2 is target) infected models (blue). Although the anomaly index $R_{t}^{*}$ may not be effective for the all-to-all attack, their RMMD statistics $R_{t}$ are much larger than those of uninfected labels in the single-target attack.}
\end{minipage}
\vspace{-8mm}
\end{figure*}

\noindent \textbf{Clean data for deviation measurement.~} 
To achieve a high accuracy in discriminating the mixed representations of benign and poisoned samples, a small set of clean samples is required for estimating the threshold in advance~\cite{chou2020sentinet,gao2019strip,tang2021demon, xu2021detecting}. The above experiments show that our method can effectively detect the dynamic attack with 30 clean images per class. Our study further demonstrates that our method can perform effectively with less clean data and even with the contaminated data. As shown in Figure~\ref{fig: clean_data_for_deviation}, we can find that even with only 8 clean images, Beatrix can still accurately identify the infected class. Moreover, we also test the robustness of Beatrix when the clean data is moderately contaminated. Figure~\ref{fig: clean_data_for_deviation_under_contamination} shows that Beatrix is still effective when no more than 16\% (or 5 images) of the clean images per class are contaminated with poisoned ones. We also compare Beatrix with the OOD detection method~\cite{sastry2020detecting} under this data contaminating scenario. Figure~\ref{fig: ood vs beatrix} shows that Beatrix is much more robust than the OOD detector when the clean data is contaminated by trojaned samples. As a result, the OOD detection method~\cite{sastry2020detecting} cannot be directly applied to the backdoor detection.

\noindent \textbf{The order of Gram matrix.~}
In the above experiments, we consider Gram matrices from the first to the ninth order. However, incorporating high-order information induces much more computational overhead. Particularly, this overhead is of vital importance for the online scenario. Thus, it is crucial to choose an appropriate 
set of orders to achieve a better trade-off between detection effectiveness and computational complexity. Specifically, we construct the detector with different values of $P$ and evaluate them in an online setting. Figure~\ref{fig: order_for_deviation} illustrates that the testing time for an input sample increases from 8.9$\times 10^{-6}$s to 78.5$\times 10^{-6}$s when the order bound increases from 1 to 9. Additionally, we find that the detection capability (false positive rate) of our method is stabilized when $P \geq 4$. 
We note that the OOD detector \cite{sastry2020detecting} needs much higher-order Gram matrices (\textit{i.e.}, $P=10$) to discriminate between in-distribution and out-of-distribution datasets, especially when the two datasets are similar-looking.
This experimental result shows that, considering the computational efficiency, it is sufficient to employ up to the third or the fourth order information to capture discriminative characteristics of benign and malicious inputs. Therefore, for the remaining experiments, we set $P$ as~4 (see efficiency comparison in Section~\ref{sec: comparison}).

\noindent \textbf{Defending against all-to-all attack.~} Here, we consider another type of adversary who launches all-to-all attacks~\cite{nguyen2020input}. Specifically, for a $c$-way classifier, the trojaned samples originally in the $i$-th class are misclassified into the (($i$+1) mod $c$)-th class. Since samples from all classes are infected by the all-to-all attack, the anomaly index values $R_{t}^{*}$ are no longer effective. However, as shown in Figure~\ref{fig: all2all}, most of the RMMD statistics $R_{t}$ of different infected labels in the all-to-all infected models are much larger than those of uninfected labels in the single-target attack. This demonstrates that Beatrix can still effectively defend against all-to-all attack relying on the RMMD statistics. In addition, we note that the all-to-all attack is the worst case of multi-target attacks that all labels are infected. However, the more labels that are infected, the less stealthy and lower performance the backdoor attack shows, especially for large datasets. As discussed in the recent research~\cite{tang2021demon}, when half of the labels are infected on the 1000-class ImageNet model, its clean accuracy and attack success rate drops 5\% and 41\%, respectively. Additionally, we consider the all-to-all attack with dynamic backdoors, which is much stronger than the all-to-all attack with universal backdoors~\cite{xu2021detecting}.

\subsection{Comparison}
\label{sec: comparison}

In this subsection, we compare our method with several state-of-the-art methods under two scenarios that are used in~\cite{tang2021demon}. In the offline protection setting, a dataset containing benign and malicious samples are processed at once, and the defender is supposed to determine whether a data class is benign or infected. On the other hand, in the online setting, samples are processed one by one, and the defender is supposed to determine whether a sample is legitimate or malicious. 

\begin{figure}[h]
\centering
\subfigure[]{
\begin{minipage}[t]{0.068\textwidth}
\centering
\includegraphics[width=0.9\textwidth]{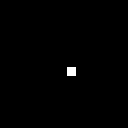}
\end{minipage}}
\subfigure[]{
\begin{minipage}[t]{0.068\textwidth}
\centering
\includegraphics[width=0.9\textwidth]{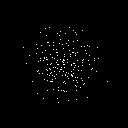}
\end{minipage}}
\subfigure[]{
\begin{minipage}[t]{0.068\textwidth}
\centering
\includegraphics[width=0.9\textwidth]{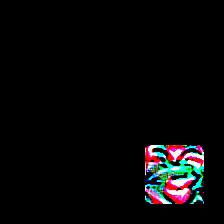}
\end{minipage}}
\subfigure[]{
\begin{minipage}[t]{0.068\textwidth}
\centering
\includegraphics[width=0.9\textwidth]{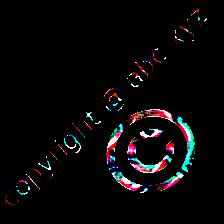}
\end{minipage}}
\caption{\label{fig: examples of universal triggers} Four static triggers used in our experiments.}
\end{figure}

In the offline setting, we launch the dynamic backdoor attack~\cite{nguyen2020input} to generate 43 infected models with respect to the 43 classes in GTSRB and 10 infected models for CIFAR10. Additionally, we launch the conventional backdoor attack~\cite{gu2019badnets}
to generate 172 (43$\times$4) infected models for GTSRB and 40 (10$\times$4) infected models for CIFAR10 with four different static triggers (Figure \ref{fig: examples of universal triggers}) which are also used  in~\cite{chou2020sentinet, gao2019strip, wang2019neural, tang2021demon}. For the online setting, we randomly select 4,000 samples from each dataset as testing samples, half of which carry dynamic (or static) triggers.

\begin{table}[t]
\centering
\caption{\label{tabel: comparison_on_dynamic_backdoor} Defense performance against dynamic backdoors on GTSRB and CIFAR10.}
\resizebox{0.45\textwidth}{!}{
\begin{tabular}{cccccccc}
\toprule
\multirow{2}{*}{\textbf{Scenario}}        & \multirow{2}{*}{\textbf{Method}} & \multicolumn{3}{c}{\textbf{GTSRB}} & \multicolumn{3}{c}{\textbf{CIFAR10}}  \\
\cmidrule(r){3-5}\cmidrule(r){6-8}
 & & \textbf{REC}(\%) & \textbf{PRE}(\%) & \textbf{F1} (\%) & \textbf{REC}(\%) & \textbf{PRE}(\%) & \textbf{F1}(\%) \\ 
\midrule
\multirow{5}{*}{Offline} 
& NC       & 18.6  & 9.3  & 12.4    & 40.0 & 25.0 & 30.8 \\
& ABS      & 27.0  & 41.5 & 32.7      & 37.0 & 49.3 & 42.3  \\
& MNTD     & 55.8 & 91.8 & 69.4     & 57.4 & 63.8 & 60.4 \\
& AC       & 74.4 & 6.0  & 11.1      & 80.0 & 28.26 & 42.1 \\
& SCAn     & 44.2 & 31.7 & 36.9     & 40.0 & 57.1 & 47.1 \\
& Beatrix     & \textbf{95.3} & \textbf{87.2} & \textbf{91.1}     & \textbf{100.0} & \textbf{83.3} & \textbf{90.9}  \\ 
\midrule
\multirow{4}{*}{Online}  
& STRIP    & 23.0 & 41.9 & 29.7     & 20.9 & 43.9 & 28.3  \\
& SentiNet & 0.00 & 0.00 & 0.00     & 0.00 & 0.00 & 0.00  \\
& SCAn     & 83.3 & 68.7 & 75.3     & 28.6 & 39.1 & 33.0  \\
& Beatrix     & \textbf{99.8} & \textbf{99.8} & \textbf{99.8}     & \textbf{99.0} & \textbf{95.4} & \textbf{97.2}  \\
\bottomrule
\end{tabular}}
\end{table}

\subsubsection{Offline defense}
In the offline setting, we consider four competitive methods, namely Neural Cleanse (NC)~\cite{wang2019neural}, ABS~\cite{liu2019abs}, MNTD~\cite{xu2021detecting}, Activation Clustering (AC)~\cite{chen2019detecting}, and SCAn~\cite{tang2021demon}. We re-implemented AC according to the paper~\cite{chen2019detecting} since the source code is not publicly available. We also re-implemented SCAn using TensorFlow Probability (TFP) based on the original MATLAB version~\cite{codescan}. Moreover, we used the PyTorch version of NC~\cite{codeinputaware}, ABS~\cite{codeabs} and MNTD~\cite{codemntd}. The comparison results are presented in Tables~\ref{tabel: comparison_on_dynamic_backdoor} and~\ref{tabel: comparison_on_conventional_backdoor}. 
The results indicate that our proposed method largely outperforms all four benchmark methods against dynamic backdoors while slightly outperforming them against universal backdoors. 

NC leverages an optimization-based reverse engineering approach to find a trigger pattern that causes any benign input from other classes to be misclassified into a target label. However, in dynamic backdoor attacks, triggers are unique and non-reusable instead of being static and universal. As a result, the reverse-engineered triggers obtained by NC are visually and functionally different from the actual dynamic triggers. 
As for the 43 trojaned models infected by the dynamic backdoor on GTSRB, there are $43\times1=43$ poisoned classes (positives) and $43\times42=1806$ benign classes (negatives). The experimental results show that NC is not effective against dynamic backdoor attacks, achieving 18.6\% (8/43) recall, and 9.3\% (8/86) precision and 12.4\% F1-score in GSTRB.

AC utilizes a two-class clustering method to separate the benign and malicious samples based on their feature vectors (activations). Specifically, AC performs dimensionality reduction using Independent Component Analysis (ICA), and then clusters them using 2-means. A high silhouette score~\cite{rousseeuw1987silhouettes} of the clustering results indicates the class is infected because the two clusters obtained by 2-means do fit the data well.  However, under the dynamic backdoor attack, the feature presentations become less distinguishable so that AC cannot effectively separate the activations of benign and trigger-carrying samples. Therefore, AC achieves 92.5\% F1-score against universal backdoor attacks whereas it yields only 11.1\% F1-score against dynamic backdoor attacks on GTSRB.

\begin{table}[t]
\centering
\caption{\label{tabel: comparison_on_conventional_backdoor} Defense performance against universal backdoors on GTSRB and CIFAR10. }
\resizebox{0.45\textwidth}{!}{
\begin{tabular}{cccccccc}
\toprule
\multirow{2}{*}{\textbf{Scenario}}        & \multirow{2}{*}{\textbf{Method}} & \multicolumn{3}{c}{\textbf{GTSRB}} & \multicolumn{3}{c}{\textbf{CIFAR10}}  \\
\cmidrule(r){3-5}\cmidrule(r){6-8}
 & & \textbf{REC}(\%) & \textbf{PRE}(\%) & \textbf{F1}(\%) & \textbf{REC}(\%) & \textbf{PRE}(\%) & \textbf{F1}(\%) \\ 
\midrule
\multirow{5}{*}{Offline}
& NC       & 97.1 & 61.6 & 75.4        & 92.5 & 51.4 & 66.7  \\ 
& ABS      & 95.3 & 81.2 & 87.7        & 90.0 & 48.6 & 63.2  \\
& MNTD     & 90.8 & 81.5 & 85.8     & 77.2 & 77.4 & 77.3 \\
& AC       & 96.5 & 88.8 & 92.5        & 87.5 & 70.6 & 79.1  \\ 
& SCAn     & 95.9 & \textbf{96.5} & 96.2        & 92.5 & 90.2 & 91.4  \\
& Beatrix     & \textbf{97.7} & 96.0 & \textbf{96.8}   & \textbf{95.0} & \textbf{90.5} & \textbf{92.7}  \\ 
\midrule
\multirow{4}{*}{Online} 
& STRIP    & 86.7 & 97.9 & 91.9        & 87.8 & 96.1 & 91.7  \\
& SentiNet & 91.5 & 96.2 & 93.8        & 90.3 & 96.9 & 93.5  \\
& SCAn     & 88.0 & 95.1 & 91.4        & 91.0 & 96.2 & 93.5  \\
& Beatrix     & \textbf{99.8} & \textbf{96.4} & \textbf{98.1}        & \textbf{97.2} & \textbf{97.1} & \textbf{97.2}  \\
\bottomrule
\end{tabular}}
\end{table}

\begin{figure*}[t]
\centering
\subfigure[Universal backdoor]{
\begin{minipage}[t]{0.22\textwidth}
\centering
\includegraphics[width=0.8\textwidth]{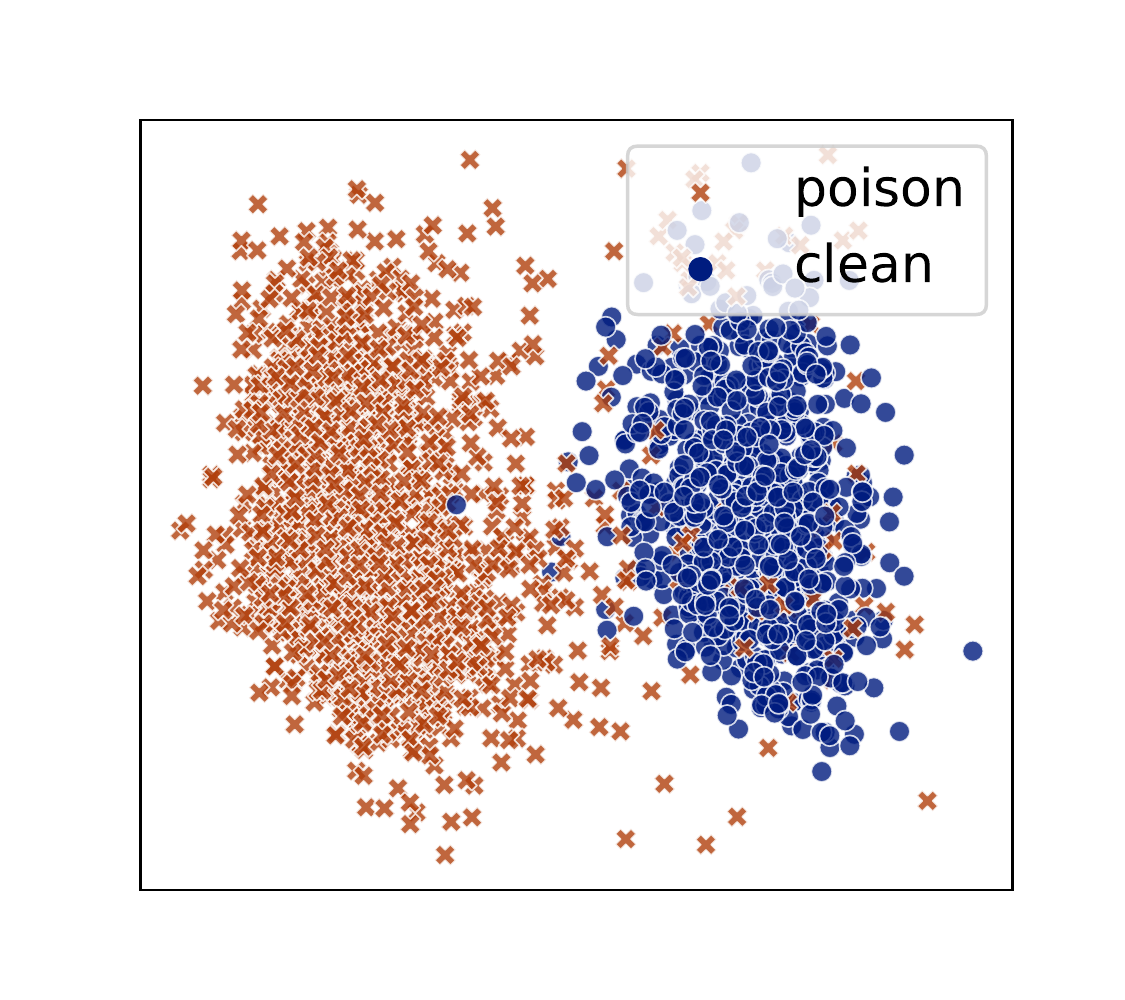}
\end{minipage}}
\subfigure[Dynamic backdoor]{
\begin{minipage}[t]{0.22\textwidth}
\centering
\includegraphics[width=0.8\textwidth]{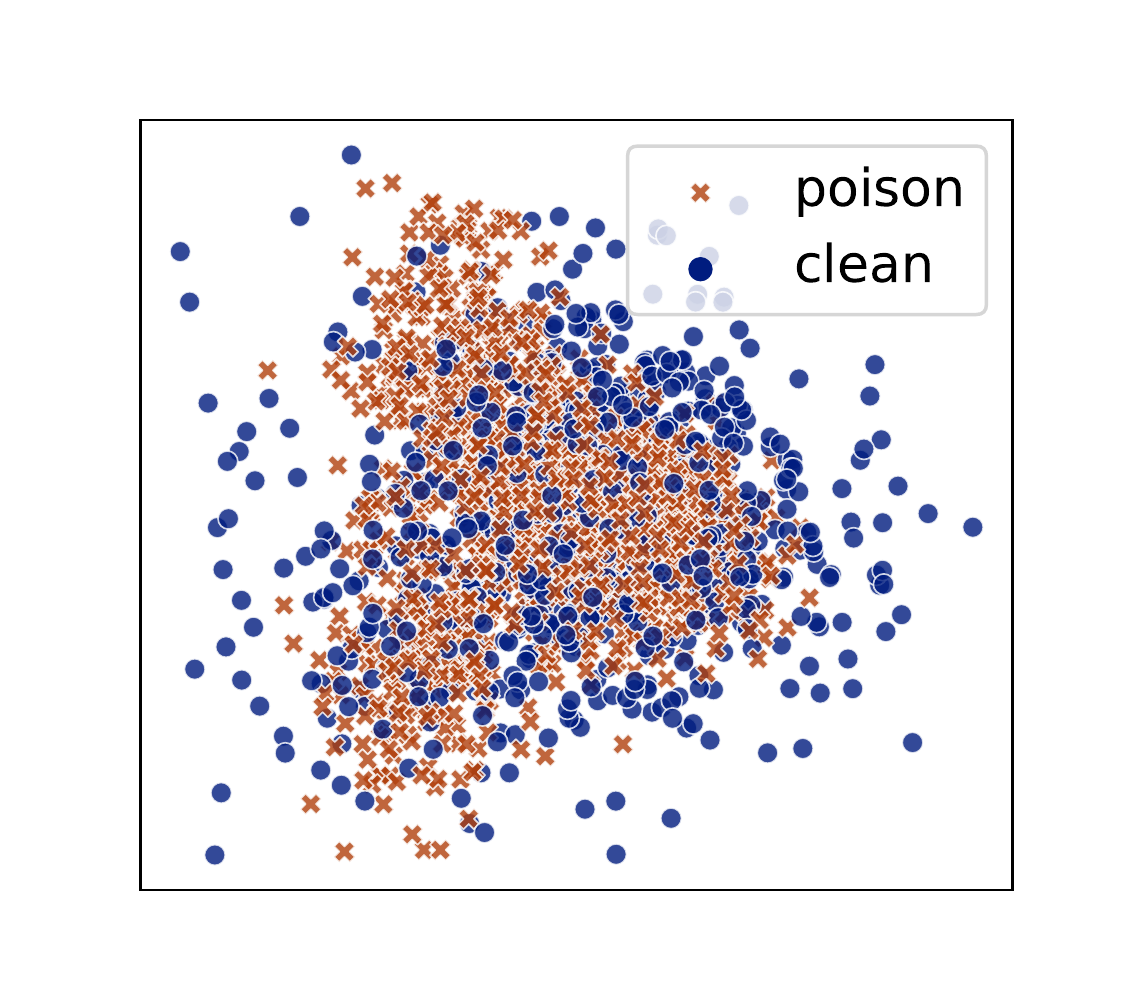}
\end{minipage}}
\subfigure[Gramian of dynamic backdoor]{
\begin{minipage}[t]{0.22\textwidth}
\centering
\includegraphics[width=0.8\textwidth]{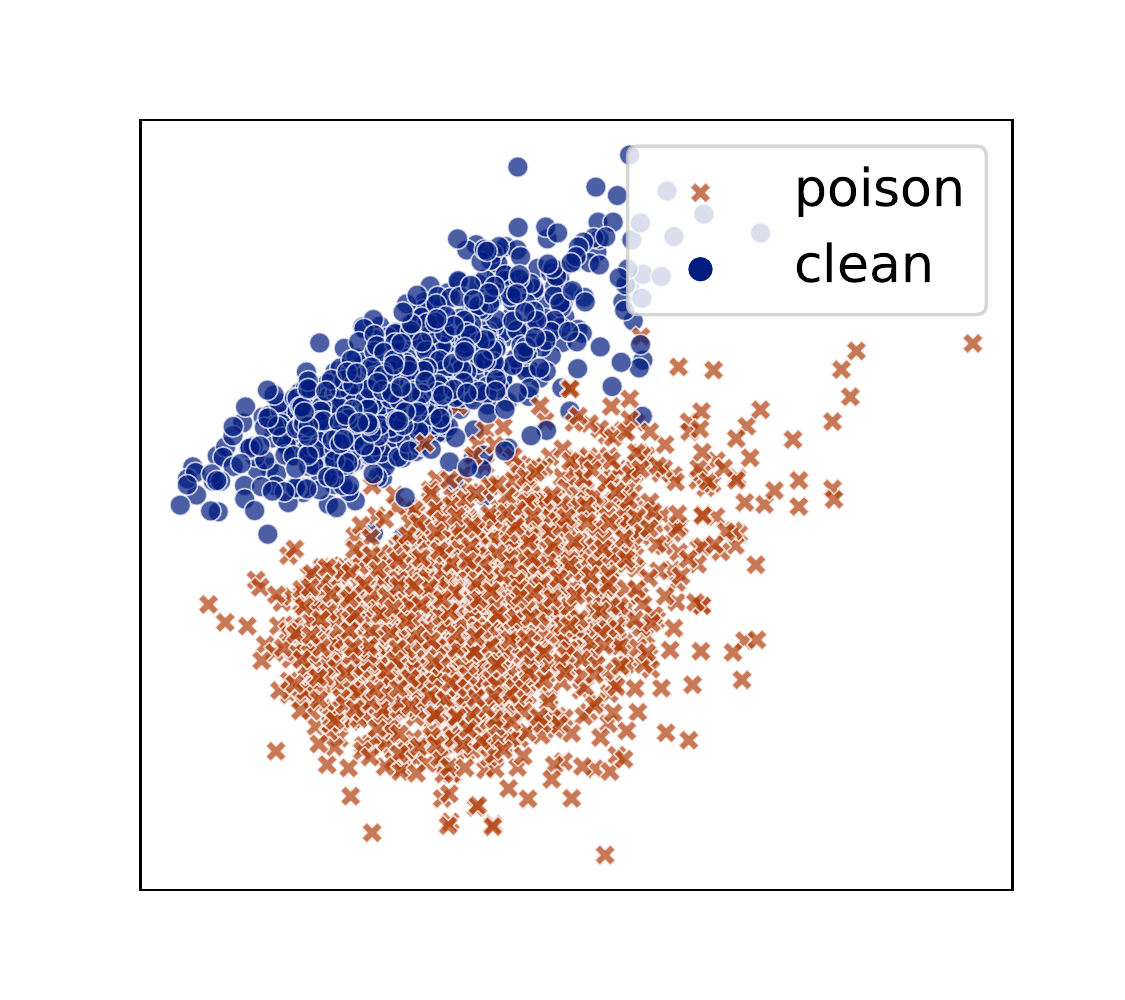}
\end{minipage}}
\subfigure[]{
\begin{minipage}[t]{0.22\textwidth}
\centering
\includegraphics[width=0.88\textwidth]{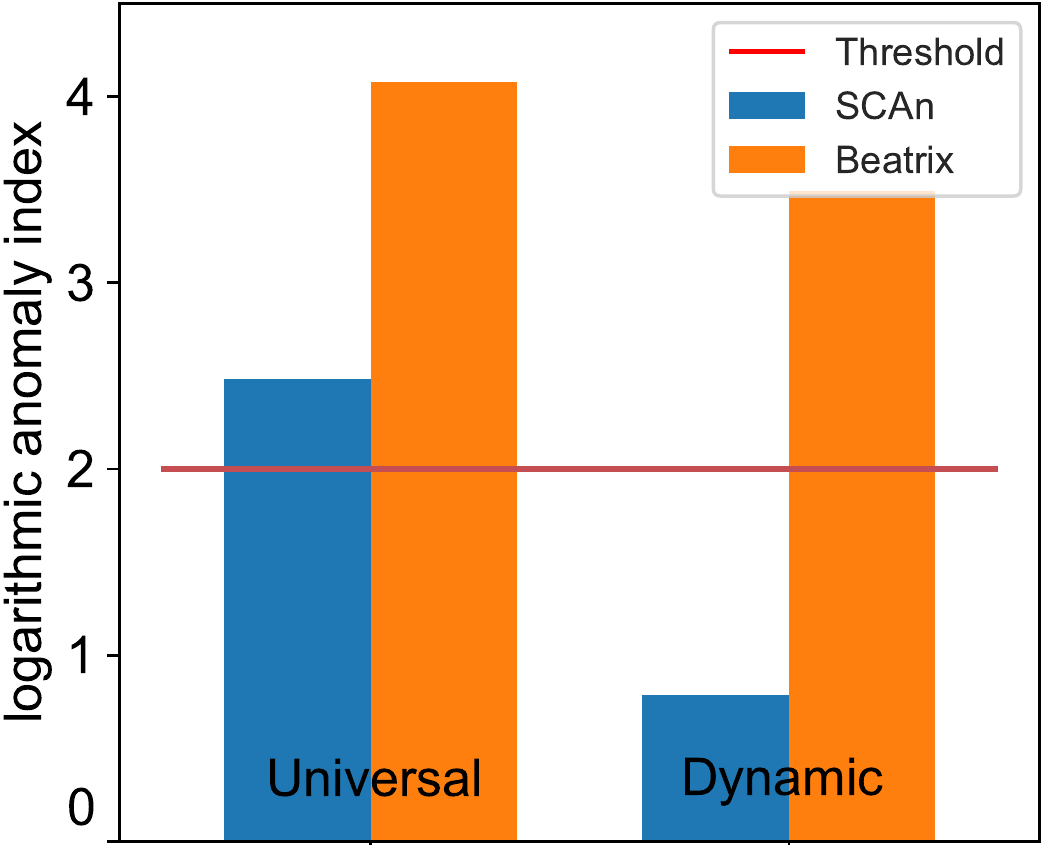}
\end{minipage}}
\caption{\label{fig: scan vs beatrix} A case study of SCAn and Beatrix on CIFAR10 with 0-th label being the trojan target. (a) and (b) illustrate the target class’ representations projected onto their first two principle components under the universal and dynamic backdoor attack, respectively. (c) shows the projection of the Gramian features $s$ of representations in (b). (d) shows the logarithmic anomaly index returned by SCAn and Beatrix.}
\vspace{-8mm}
\end{figure*}

SCAn models the representation distribution by a Gaussian distribution so that it uses a set of clean samples to estimate the covariance matrix. Then SCAn leverages Linear Discriminant Analysis (LDA) to separate the feature representations into two subgroups. A high statistic value from the likelihood-ratio test indicates the class is infected. As we discussed in Section~\ref{sec: Limitations against Dynamic Backdoors}, the mean discrepancy is ineffective and the universal covariance assumption is not true in the dynamic backdoor attack. As a result, SCAn is ineffective against the dynamic backdoor attack because of its intrinsic limitations. 
Its F1-score against universal attacks on GSTRB is 96.2\%,  but it drops to 36.9\% when it encounters dynamic attacks. 
Figure~\ref{fig: scan vs beatrix} provides a more in-depth analysis from the representation space perspective on CIFAR10. It shows that SCAn cannot raise the alarm when the feature representations of clean and poisoned samples are deeply fused in the dynamic backdoor. In contrast, Beatrix remains effective against the dynamic backdoor since Gram matrix is capable of capturing the subtle differences between clean and poisoned representations, as shown in Figure~\ref{fig: scan vs beatrix}(c).

ABS can only determine whether a model is infected or not. Therefore, we trained 100 clean models and 100 infected models with random initialization for the evaluations in universal and dynamic attacks. ABS assumes that each target label is associated with only one trigger and the trigger subverts all benign samples to the target label. This assumption is broken by the dynamic backdoor attack, in which each trojaned sample has its own unique trigger. 
Additionally, based upon the universal trigger assumption, ABS assumes there is only one compromised neuron activated at a time by the trigger. Put differently, the changes in the activation of the intermediate layer only depend on this single neuron when encountering a trigger-carrying sample. However, due to the uniqueness of dynamic triggers, the abnormal changes in the activation pattern are dispersed across multiple neurons.
Therefore, their reverse engineered trigger cannot reflect the malfunction in the model infected by dynamic backdoor attacks, and consequently, it yields only 32.7\% and 42.3\% F1-score on GSTRB and CIFAR10 against dynamic backdoor attacks. 

MNTD, similar to ABS, can only flag a model as either trojaned or benign. Therefore, we evaluate MNTD with 100 clean/infected models used in the evaluation of ABS. MNTD uses jumbo learning to generate thousands of shadow models and then train a meta-classifier to learn the output differences of trojaned between clean models. MNTD fails to detect infected models with dynamic backdoor as the representations are already deep fused in the middle activation layers (see Figure~\ref{fig: scan vs beatrix}(b)). Therefore, MNTD achieves 85.8\% F1-score against universal backdoor attacks on GTSRB whereas it drops to 69.4\% when it encounters dynamic attacks.

\subsubsection{Online defense}

In the online setting, we consider three existing defenses, STRIP~\cite{gao2019strip}, SentiNet~\cite{chou2020sentinet}, and SCAn~\cite{tang2021demon}. We re-implemented SentiNet according to the paper, and used the Pytorch version of STRIP~\cite{codeinputaware}. SCAn is configured for online defense following the paper. The experimental results are shown in Tables~\ref{tabel: comparison_on_dynamic_backdoor} and~\ref{tabel: comparison_on_conventional_backdoor}.

STRIP works by superimposing a set of randomly selected clean images to an input image and measuring the entropy of the prediction outputs. In the conventional backdoor attack, a trojaned image with a static trigger is resistant to this perturbation, leading to a much lower entropy of the outputs compared to that of a benign input. The effectiveness of STRIP relies on the dominant impact of the trigger~\cite{tang2021demon}. However, this assumption no longer holds in the dynamic backdoor attack where images with mismatching triggers will deactivate the backdoors~\cite{nguyen2020input, li2021invisible}, and subsequently the F1-score of STRIP drops from 91.9\% and 91.7\% against universal backdoors to only 29.7\% and 28.3\% against dynamic backdoors.

\begin{table}[t]
\centering
\caption{\label{table: efficiency comparison} Efficiency comparison.}
\resizebox{0.4\textwidth}{!}{
\begin{tabular}{ccccc} 
\toprule
\textbf{Method} & STRIP & SentiNet & SCAn & Beatrix  \\ 
\midrule
\textbf{Time} (s)    & 0.04 & 0.11 & 13.58 & 35.14$\times10^{-6}$  \\
\bottomrule
\end{tabular}}
\end{table}

SentiNet utilizes the model interpretability technique~\cite{selvaraju2017grad} to locate highly salient contiguous region (\textit{i.e.}, a potential trigger-region) of a given input. The extracted salient region is then overlaid on a set of clean images whose classification results are used to distinguish normal images from malicious images, since the trigger region is much more likely than normal region to subvert the clean images to the target label. What has been assumed here is that the attack is \textit{localized} (\textit{i.e.}, the trigger-region is constrained to a small contiguous part) and \textit{universal} (\textit{i.e.}, the attack is sample-agnostic). Both assumptions are broken by the dynamic backdoor attack where triggers are sample-specific and are distributed over different and disjointed portions of an image. Thus, the recall, precision and F1-score of SentiNet are 0\%, indicating that it is no longer effective against this advanced attack.

SCAn builds a composition model (covariance estimation) as well as an untangling model (mean estimation for each class)  on a set of clean samples in an offline manner. Therefore, for each input sample, SCAn needs to update the untangling model for the image class. 
The incoming sample is tagged as malicious if it results in a class anomaly index larger than the threshold ($e^{2}$) and falls into the smaller subgroup. 
As we discussed above, SCAn models the feature representations under the LDA assumption (see  Section~\ref{sec: Limitations against Dynamic Backdoors}), which is violated in the dynamic backdoor attack. Consequently, the error in the composition model leads to the ineffectiveness of the untangling model in distinguishing representations of benign inputs from those of malicious inputs. 
As a result, the F1-score of SCAn drops from 93.5\% (in universal attacks) to 33.5\% (in dynamic attacks).
Additionally, due to \textit{its dependency on the accumulation of adversarial inputs}~\cite{tang2021demon}, SCAn shows a suboptimal performance against universal attacks in the online setting while it achieves similar performance in the offline setting compared to Beatrix.

\noindent \textbf{Efficiency comparison.~} As the overhead is important for the online scenario, we also compare the efficiency of Beatrix with those of other online defenses on the GTSRB dataset. 
The experiment is conducted on one NVIDIA GeForce RTX 3090 GPU. The running time is averaged over 1,000 testing samples. As shown in Table~\ref{table: efficiency comparison}, our approach is much faster than the three baseline methods. SCAn consumes the longest time (13.58s) compared to other three methods since it needs to update the untangling model when there is an incoming sample. SentiNet pastes the potential trigger region of the input image on clean and noise images while STRIP directly superimposes each input on clean images. Thus, SentiNet (0.11s) is slightly  slower than STRIP (0.04s). 
The main computation overhead of Beatrix is building the deviation measurement model with clean samples, but this measurement model can be obtained through an offline training. Therefore, Beatrix only needs a forward pass to get the intermediate presentation for each sample and computes Gramian features of different orders, which can be obtained simultaneously when the user (defender) trains or tests her/his model with a dataset.
Therefore, Beatrix takes only 35.14 $\times10^{-6}$s (for $P=4$). Furthermore, as demonstrated in Section~\ref{sec: Effectiveness Against Dynamic Backdoor}, the defender can choose a different number of orders to trade off efficiency and effectiveness. 

\begin{figure}[t]
\centering
\subfigure[]{
\begin{minipage}[t]{0.068\textwidth}
\centering
\includegraphics[width=0.9\textwidth]{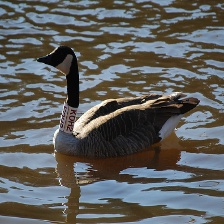}
\end{minipage}}
\subfigure[]{
\begin{minipage}[t]{0.068\textwidth}
\centering
\includegraphics[width=0.9\textwidth]{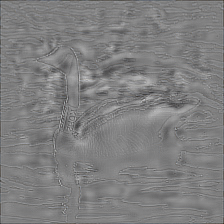}
\end{minipage}}
\subfigure[]{
\begin{minipage}[t]{0.068\textwidth}
\centering
\includegraphics[width=0.9\textwidth]{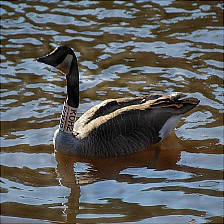}
\end{minipage}}
\subfigure[]{
\begin{minipage}[t]{0.068\textwidth}
\centering
\includegraphics[width=0.9\textwidth]{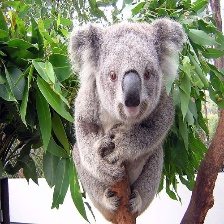}
\end{minipage}}
\subfigure[]{
\begin{minipage}[t]{0.068\textwidth}
\centering
\includegraphics[width=0.9\textwidth]{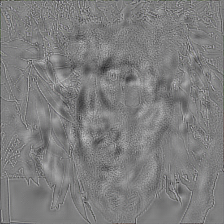}
\end{minipage}}
\subfigure[]{
\begin{minipage}[t]{0.068\textwidth}
\centering
\includegraphics[width=0.9\textwidth]{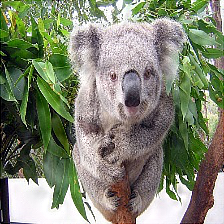}
\end{minipage}}
\caption{\label{fig: examples of issba} Examples of poisoned samples under ISSBA. (a) and (d) are clean images, (b) and (e) are sample-specific triggers, and (c) and (f) are poisoned images.}
\vspace{-6mm}
\end{figure}

\begin{figure}
\centering
\includegraphics[width=0.95\linewidth]{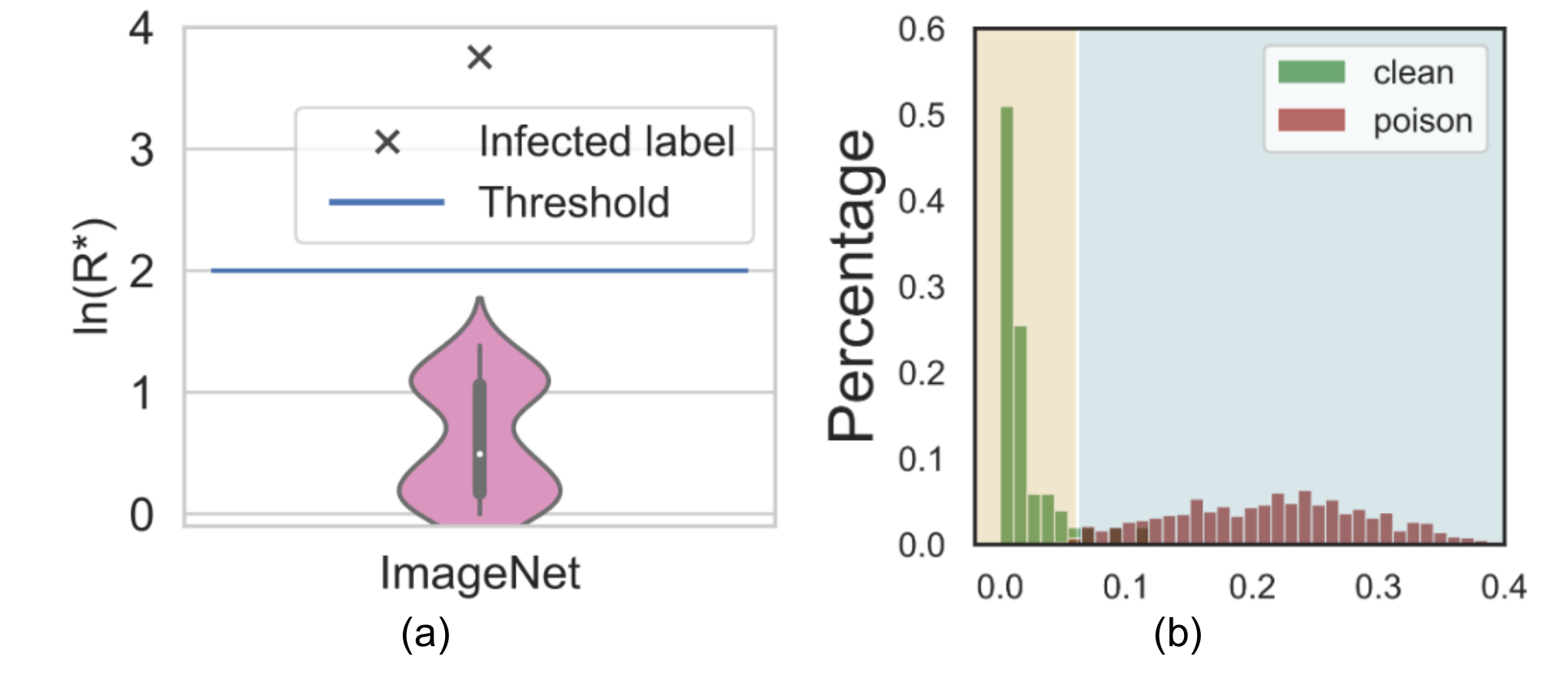}
\caption{(a) The logarithmic anomaly index of infected and uninfected labels under ISSBA. (b) Deviation distribution of benign and trojaned samples in the infected class under ISSBA.}
\label{fig: defend_against_issb}
\end{figure}

\subsection{Robustness Against Other Attacks}
\label{sec: Robustness Against Other Attacks}

\begin{figure}[t]
\centering
\subfigure[]{
\begin{minipage}[t]{0.068\textwidth}
\centering
\includegraphics[width=0.9\textwidth]{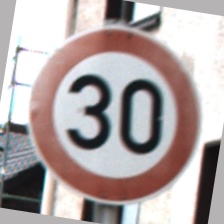}
\end{minipage}}
\subfigure[]{
\begin{minipage}[t]{0.068\textwidth}
\centering
\includegraphics[width=0.9\textwidth]{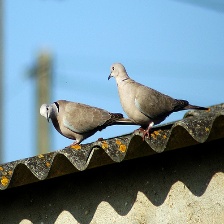}
\end{minipage}}
\subfigure[]{
\begin{minipage}[t]{0.068\textwidth}
\centering
\includegraphics[width=0.9\textwidth]{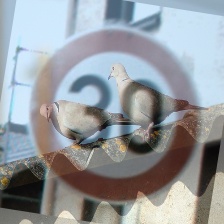}
\end{minipage}}
\subfigure[]{
\begin{minipage}[t]{0.068\textwidth}
\centering
\includegraphics[width=0.9\textwidth]{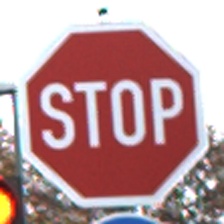}
\end{minipage}}
\subfigure[]{
\begin{minipage}[t]{0.068\textwidth}
\centering
\includegraphics[width=0.9\textwidth]{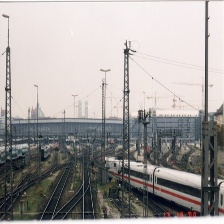}
\end{minipage}}
\subfigure[]{
\begin{minipage}[t]{0.068\textwidth}
\centering
\includegraphics[width=0.9\textwidth]{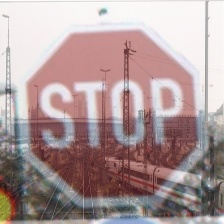}
\end{minipage}}
\caption{\label{fig: examples of refool} Examples of poisoned samples under \textit{Refool}. (a) and (d) are clean images, (b) and (e) are reflection patterns, and (c) and (f) are poisoned images.}
\vspace{-6mm}
\end{figure}

\begin{figure}
\centering
\includegraphics[width=0.95\linewidth]{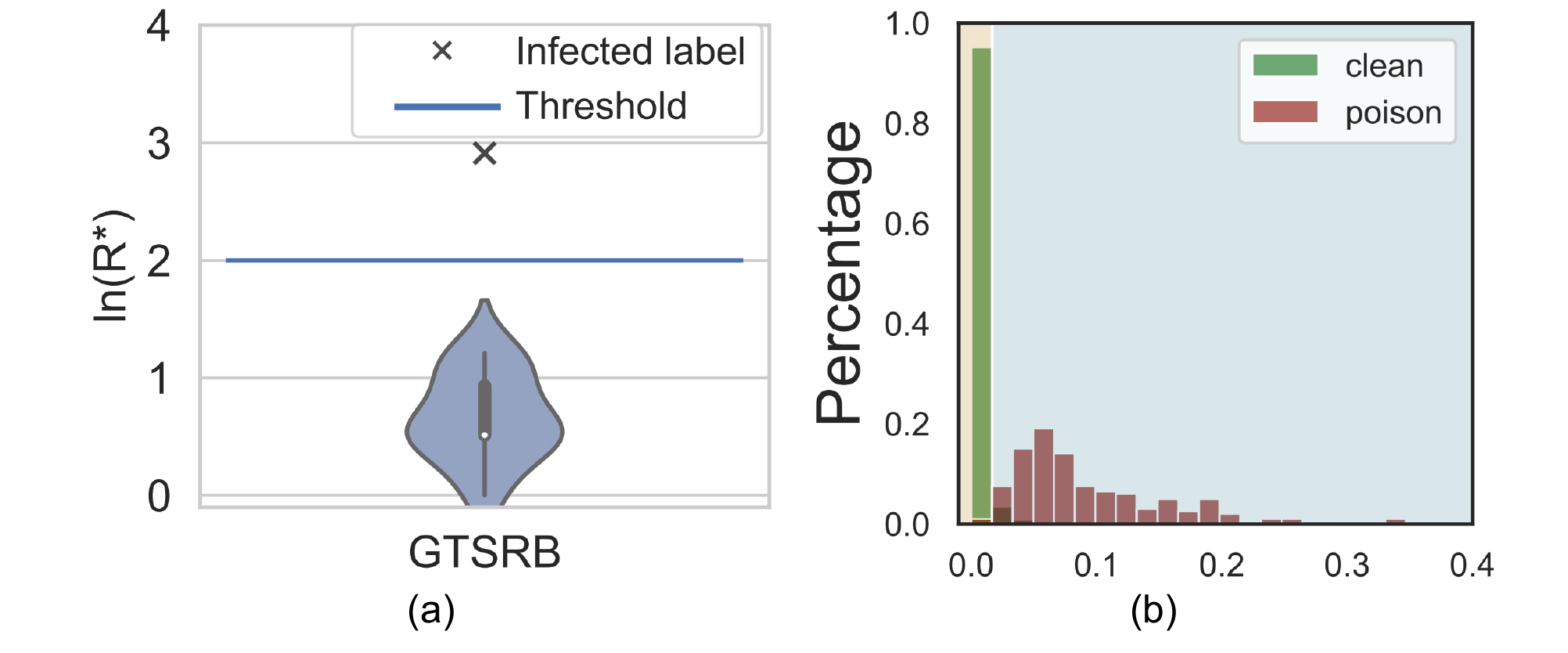}
\caption{(a) The logarithmic anomaly index of infected and uninfected labels under \textit{Refool}. (b) Deviation distribution of benign and trojaned samples in the infected class under \textit{Refool}.}
\label{fig: defend_against_refool}
\end{figure}

\noindent \textbf{Invisible sample-specific backdoor attack.~}
Motivated by the advances in DNN-based image steganography~\cite{tancik2020stegastamp, baluja2017hiding}, Li \textit{et al.} proposed an invisible sample-specific backdoor attack (ISSBA),  where triggers are generated by a pre-trained encoder network~\cite{li2021invisible}. The generated triggers are invisible additive noise containing the information of a representative string of the target label. The attacker can flexibly design the string as the name and the index of the target class or even a random character. The encoder network embeds the string into a clean image to obtain a poisoned image. Thus, the poisoned image generator (encoder) is conditioned on the input images, indicating the backdoor triggers vary from input to input. A DNN classifier trained on the poisoned images will misclassify images trojaned by the same encoder into the classes indicated by the embedded strings. Some examples of the poisoned ImageNet samples and their corresponding triggers are shown in Figure~\ref{fig: examples of issba}. 

We evaluate Beatrix on this attack using the infected model and the dataset shared by the authors~\cite{li2021invisible}. 
Since they only release their implementation on the ImageNet dataset, we use the released infected model trained on a ImageNet subset which  contains 200 classes with the 0-th label (goldfish) being the trojan target.
As shown in Figure~\ref{fig: defend_against_issb}, the anomaly index of the infected label (label~0) is much larger than the uninfected labels (labels 1-199), indicating that Beatrix can effectively defend against this attack. More detailed comparison with other detection methods can be found in Appendix~\ref{sec: Defense Performance Against ISSBA}.

\noindent \textbf{Reflection backdoor attack.~} Liu \textit{et al.} proposed \textit{reflection backdoor} (\textit{Refool}) using a special backdoor pattern based on a nature phenomenon --- reflection~\cite{liu2020reflection}. Reflection occurs wherever there are glasses or smooth surfaces. \textit{Refool} generates poisoned images by adding reflections to clean images based on mathematical models of physical reflection scenarios. Different from conventional backdoor attacks that rely on a fixed trigger pattern, \textit{Refool} can utilize various reflections as the trigger pattern, making it stealthier than other attacks. 
Additionally, \textit{Refool} blends the clean image with a triggering reflection pattern, so that the trigger is complex and spans all over the image.
Some examples of clean images and their poisoned counterparts with reflection triggers are illustrated in Figure~\ref{fig: examples of refool}.

We evaluate Beatrix against this attack using the code and datasets shared by the authors~\cite{liu2020reflection}. The released datasets include three traffic sign datasets: GTSRB, BelgiumTSC~\cite{timofte2014multi} and CTSRD~\cite{CTSRD}. 
In this experiment, we use the GTSRB dataset and randomly choose the reflection images from PascalVOC~\cite{everingham2011pascal} following the original implementation.
The target class is the speed limit sign of 30 km/h. Our detection results are demonstrated in Figure~\ref{fig: defend_against_refool}. It shows that the anomaly index of the infected label is larger than the threshold, and those of the uninfected labels are all below the threshold. For online detection, Beatrix can effectively distinguish clean images from trigger-carrying ones with 99.99\% TPR @ 5\% FPR and 98.50\% TPR @ 1\% FPR, as shown in Figure~\ref{fig: defend_against_refool}(b).

\begin{figure}[t]
\centering
\includegraphics[width=0.9\linewidth]{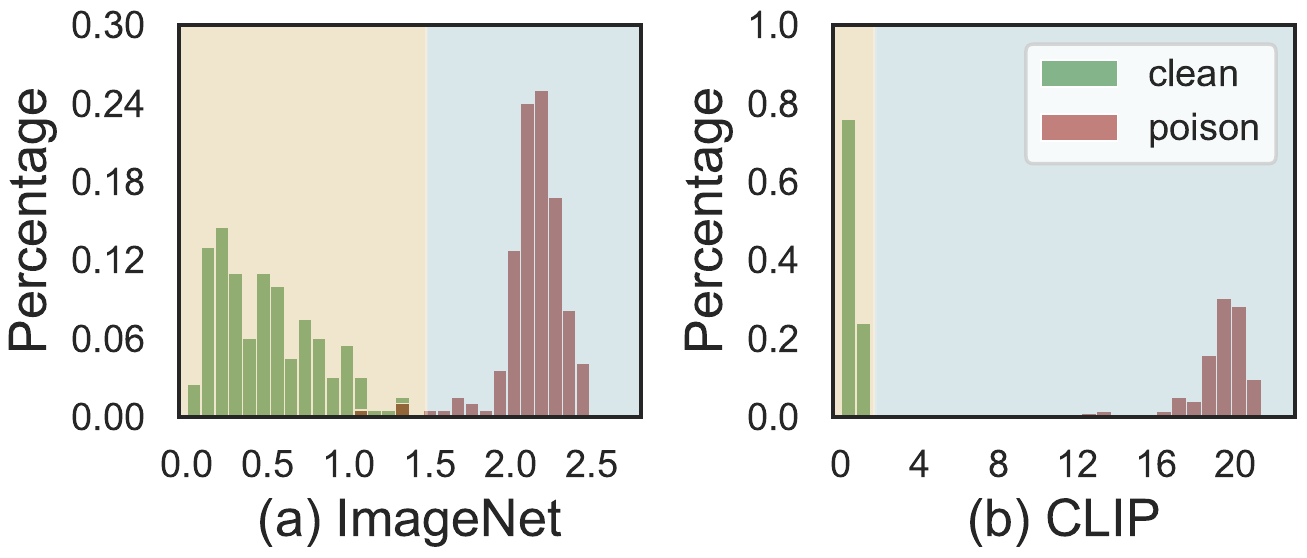}
\caption{\label{fig: defend_against_badencoder} Deviation distribution of benign and trojaned samples in the infected class of (a) Imagnet encoder and (b) CLIP encoder under BadEncoder attack.}
\end{figure}

\noindent \textbf{BadEncoder attack.~} Another recently introduced backdoor attack is BadEncoder~\cite{jia2022badencoder}, which has been proposed for self-supervised learning pipelines. Self-supervised learning aims to pre-train an image encoder using a large amount of unlabeled data. Thus, the pre-trained encoder can be used as a feature extractor to build downstream classifiers in many different tasks. BadEncoder aims to compromise the self-supervised learning pipeline by injecting backdoors into a pre-trained image encoder such that the downstream classifier built upon this trojaned encoder will inherit the backdoor behavior. 
To craft a trojaned image encoder, BadEncoder fine-tunes a clean image encoder with two additional loss terms named effectiveness loss and utility loss. The effectiveness loss measures the similarity between feature vectors of reference inputs (\textit{i.e., clean inputs in the target class}) and those of trigger-carrying inputs produced by the trojaned encoder. 
To maintain utility as well as stealthiness, BadEncoder applies the utility loss to encourage the trojaned encoder and the clean encoder produces similar outputs given the same clean inputs. In this way, a downstream classifier built upon the trojaned encoder will behave normally on clean inputs but misclassify trigger-carrying inputs into the target class since their feature representations are similar to those of clean target inputs. 

We put Beatrix into test against BadEncoder using the infected models and datasets published along with the paper that introduced the attack itself~\cite{jia2022badencoder}. In the experiments, we consider two real-world image encoders: ImageNet encoder originally released by Google~\cite{chen2020simple} and CLIP encoder by OpenAI~\cite{radford2021learning}. The target downstream dataset is GTSRB, and the target class is the 12th-label (\textit{i.e.}, the priority road sign). Instead of building a downstream classifier, we directly evaluate Beatrix on distinguishing clean and poison inputs based on their feature vectors produced by the backdoor encoder.
As shown in Figure~\ref{fig: defend_against_badencoder},  Beatrix can effectively defend against BadEncoder. Specifically, Beatrix achieves 99.8\% TPR @ 1\% FPR on the infected ImageNet encoder, and 100\% TPR @ 1\% FPR on the infected CLIP encoder.

\subsection{Beyond the Image Domain}

\begin{figure}[t]
\centering
\includegraphics[width=0.9\linewidth]{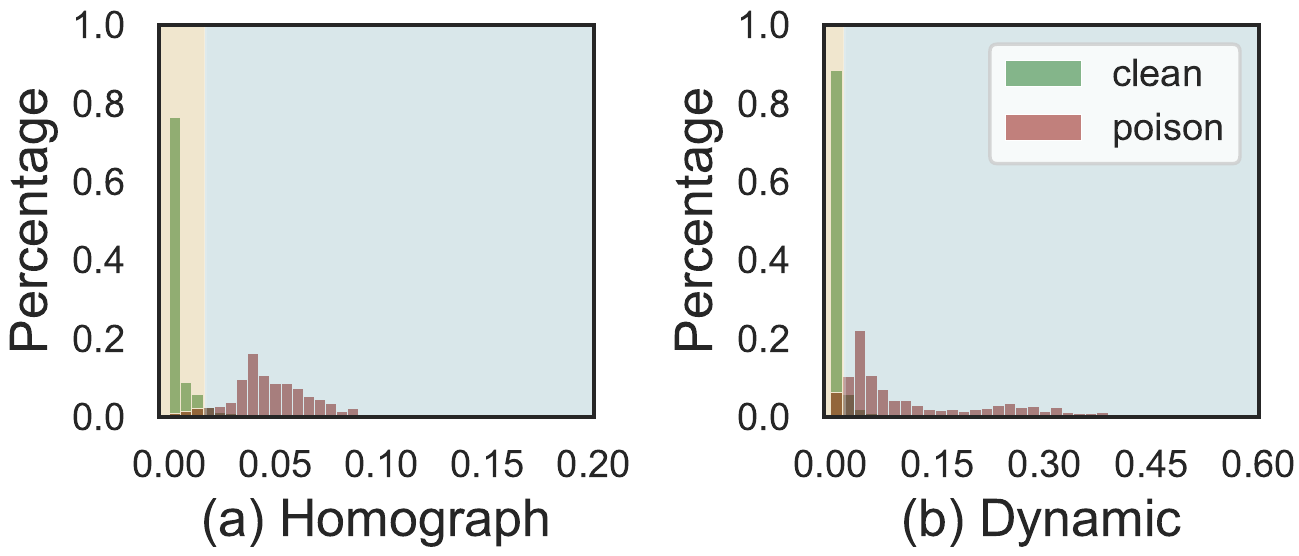}
\caption{\label{fig: defend_against_nlp} Deviation distribution of benign and trojaned samples in the infected class under (a) Homograph Backdoor Attack and (b) Dynamic Sentence Backdoor Attack.}
\vspace{-6mm}
\end{figure}

\begin{figure}[t]
\centering
\includegraphics[width=0.95\linewidth]{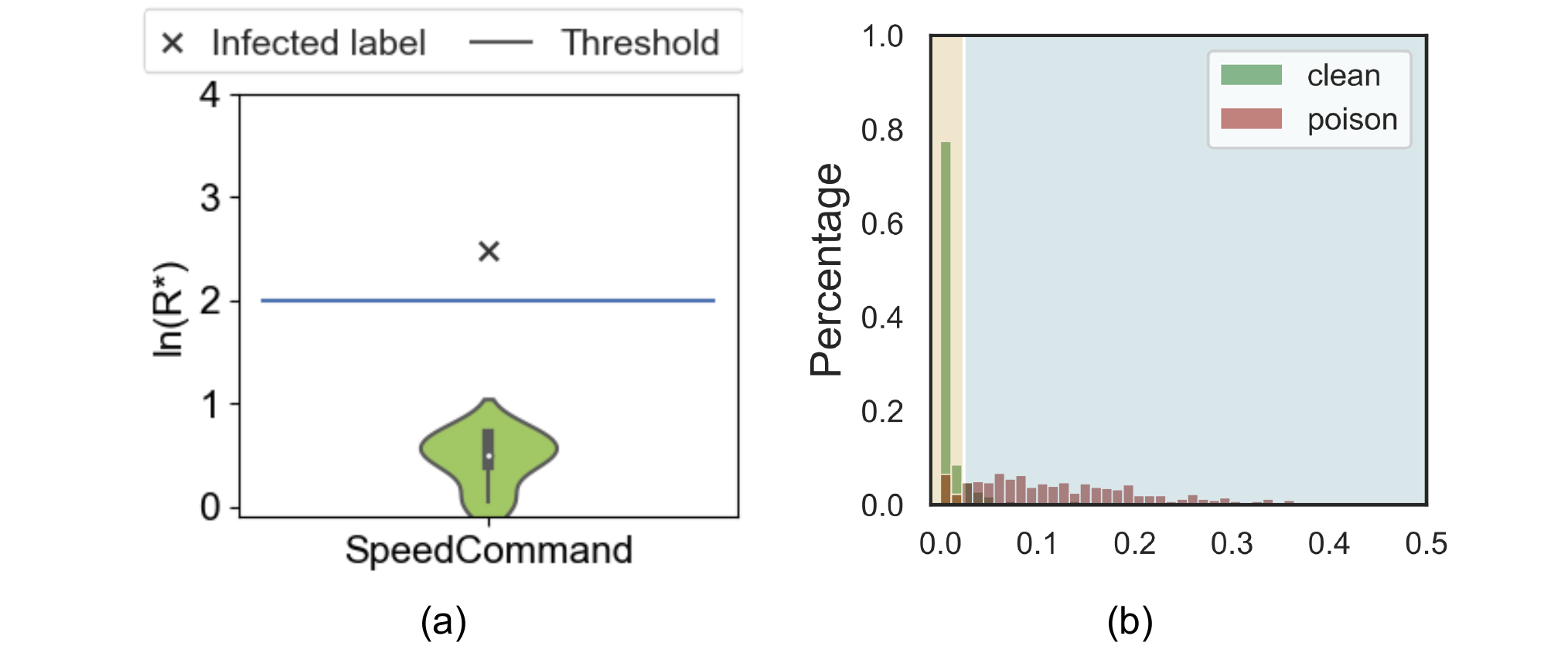}
\caption{(a) The logarithmic anomaly index of infected and uninfected labels of a speech recognition backdoor model. (b) Deviation distribution of benign and trojaned samples in the infected class under the speech recognition backdoor attack.}
\label{fig: defend_against_audio}
\end{figure}

Akin to previous works, we mainly focus on the
image classification tasks. There are backdoor attacks in other domains, such as natural language processing (NLP) \cite{li2021hidden, azizi2021t, chen2021badnl}, acoustics signal processing \cite{zhai2021backdoor} and malware detection \cite{li2021backdoor, severi2021explanation}. Here, we extend our approach to mitigate the threats posed by backdoor attacks in speech recognition and text classification domains. 

For the NLP task, we evaluate Beatrix on the homograph backdoor attack and dynamic sentence backdoor attack proposed by Li \textit{et al.} \cite{li2021hidden}. 
The homograph backdoor attack inserts triggers by replacing several characters of the clean sequences with their homograph equivalent. Given an original sentence, the dynamic sentence backdoor attack uses pre-trained language models to generate a suffix sentence to act as the trigger.
We use the code and dataset shared by the authors to train poisoned BERT model on the toxic comment classification dataset \cite{toxic2020}. To balance the number of positive (\textit{i.e.}, toxic) and negative (\textit{i.e.}, non-toxic) samples, it draws 16225 negative samples from the negative texts so the final dataset consists of 32450 samples. The dataset is then split to give 29205 (90\% of the dataset) in the training set and 3245 (10\%) in the test set.
Since this is a binary classification task, we directly evaluate the online defense performance of Beatrix. As shown in Figure \ref{fig: defend_against_nlp}, Beatrix achieves 89.8\% TPR @ 5\% FPR on homograph attack and 75.2\% TPR @ 5\% FPR on dynamic sentence attack.

For the speech task, we use the speech recognition
backdoor implementation provided in the work \cite{xu2021detecting} on the SpeechCommand dataset \cite{warden2018speech}. The original dataset consists of 65,000 one-second audio files of 35 classes. Following the previous work \cite{xu2021detecting}, we use the files of 10 classes (\textit{i.e.}, “yes”, “no”, “up”, “down”, “left”, “right”, “on”, “off”, “stop”, “go”), which gives 30,769 training samples
and 4,074 testing samples. It first extracts the mel-spectrogram of each file and then trains an LSTM model over the mel-spectrograms. The backdoor trigger is a consecutive noise signal whose length is 0.05 seconds. Our detection results shown in Figure~\ref{fig: defend_against_audio} demonstrate that Beatrix can detect the infected label effectively and distinguish clean signal from trigger-carrying ones with 77.5\% TPR @ 5\% FPR.

\subsection{Adaptive Attack}
We study an adaptive adversary who targets at the deviation measurement of Beatrix. We consider a strong white-box adversary who controls the training process of the victim model. The objective of the adversary is to force the activation patterns of poisoned images to resemble those of clean images, so that Beatrix cannot separate the benign and malicious inputs. To achieve this goal, we design an adaptive loss to minimize the distance between poisoned and clean images of a target class, in the representation space based on multiple high-order Gram matrices. In the experiment, we set the order upper bound $P$ as 4 in the adaptive loss. Therefore, the loss function of the adaptive attack can be denoted as follows:
\begin{align*}
L &= L_{o} + \lambda L_{a}, \\
L_{a} = \mathbb{E}_{x\in X_{/y_t}, x_t \in X_{y_t}}&\left[\sum^{P}_{p=1}\Vert G^{p}(\mathcal{B}(x,g(x))) - G^{p}(x_{t})\Vert^{2}\right],
\end{align*}
where $L_{o}$ is the original dynamic backdoor loss for the victim model $f$ ~\cite{nguyen2020input}, and $L_{a}$ denotes the adaptive loss. $G^{p}$ denotes the $p$-th order Gram matrix of the internal activation from an input image. $X_{y_t}$ and $X_{/y_t}$ denotes target and non-target training data, respectively. $\lambda$ is a hyperparameter balancing the model performance and the adaptive strength.

We perform this attack on the GTSRB and CIFAR10 datasets. We evaluate the online performance of Beatrix to determine whether an input image is benign or malicious during the inference phase. In the experiment, we randomly select 500 clean images from the target class and 500 poisoned images from other classes. 
The results are shown in Table~\ref{tabel: adaptive attack}. We can find that the true positive rate of Beatrix slightly decreases when the hyper-parameter $\lambda$ increases from 0.05 to 0.5. In particular, when $\lambda$ increases to 1, Beatrix yields only 14.0\% TPR @ 1\% FPR on CIFAR10 dataset, indicating that Beatrix is no longer effective. However, the model performance (\textit{i.e.}, clean accuracy on benign inputs and attack success rate on  malicious ones) also decreases significantly. 

\begin{table}[t]
\centering
\caption{\label{tabel: adaptive attack} Adaptive attack.}
\resizebox{0.45\textwidth}{!}{
\begin{tabular}{ccccccc} 
\toprule
& $\lambda$& 0.05    & 0.1    & 0.5    &  1    &   5     \\ 
\midrule
\multirow{5}{*}{GTSRB} 
&CA      &  98.5\% & 98.1\% & 93.8\% & 79.7\% &   4.7\%      \\
&ASR     &  97.5\% & 96.8\% & 95.2\% & 89.1\% &   -     \\
&5\% FPR &  99.8\% & 98.6\% & 97.4\% & 92.4\% &   -     \\
&1\% FPR &  98.8\% & 96.6\% & 93.6\% & 81.0\% &   -     \\
\midrule
\multirow{5}{*}{CIFAR10} 
&CA      &  92.7\% & 91.8\% & 89.8\% & 69.5\% &   10.0\%      \\
&ASR     &  98.1\% & 96.2\% & 94.6\% & 85.1\% &   -     \\
&5\% FPR &  97.6\% & 94.3\% & 83.3\% & 40.5\% &   -     \\
&1\% FPR &  87.2\% & 76.6\% & 51.7\% & 14.0\% &   -     \\
\bottomrule
\end{tabular}}
\end{table}

\section{Related Work}
\label{sec: related work}

\noindent \textbf{Backdoor attacks.} 
With the increasing use of pre-trained DNNs in security-sensitive domains, backdoor attacks have recently been recognized as a new threat. Current attacks inject backdoors to DNNs by either poisoning the training dataset or directly manipulating the model parameters.
BadNets was first introduced as a poisoning-based backdoor attack in which the adversary has full control over the model and training data~\cite{gu2019badnets}. Subsequently, a series of attempts were made to launch attacks with fewer poisoned samples and limited access to training data~\cite{chen2017targeted, liu2018trojaning}. On the other end, to make the poisoning data stealthier, there are attacks that inject poisoned samples without altering their labels~\cite{shafahi2018poison,turner2019label}. 
Recently, backdoors activated by semantic triggers with diverse patterns were studied to further conceal the attacks~\cite{barni2019new,liu2020reflection,li2020invisible,saha2020hidden}. The threat of neural backdoors in emerging machine learning paradigms has also been investigated~\cite{yao2019latent, shafahi2018poison,xie2019dba,jia2022badencoder}. Interestingly, model replacement has been identified as an attack surface in federated learning~\cite{bagdasaryan2020backdoor}. 
Although various backdoor attacks are proposed, most of them employ static triggers which can be easily detected by existing defensive techniques. On the contrary, recently proposed dynamic backdoor attacks craft input-aware triggers, which hardens the detection of such backdoors~\cite{salem2020dynamic, nguyen2020input, li2021invisible}.

\noindent \textbf{Defenses against backdoor attacks.} 
Current defensive techniques can be broadly categorized into two branches. The first line of works conducts a model analysis to identify backdoors. Otherwise, defenders can perform run-time detection against triggered backdoors.
A common wisdom in model analysis is to discover anomalies in the learned representations~\cite{tran2018spectral, chen2019detecting}. Moreover, other methods are reverse-engineering possible triggers and unlearning inserted backdoors~\cite{wang2019neural, liu2019abs, liu2021ex}. However, these methods are reported as ineffective against large triggers and source-label-specific attacks~\cite{liu2019abs, tang2021demon}. There is also a method that adopts a meta classifier trained on a set of clean and trojaned models to identify compromised models~\cite{xu2021detecting}.
A defender can also implement run-time detection methods when the DNN has been deployed. STRIP was proposed to detect backdoors controlled by dominant triggers~\cite{gao2019strip}. SentiNet adopts Grad-Cam to detect potential backdoor locations and malicious inputs~\cite{selvaraju2017grad, chou2020sentinet}. A similar idea is also explored and extended in Februus~\cite{doan2020februus}. However, the aforementioned methods are vulnerable to adaptive attacks such as the source-label-specific backdoor. Lately, SCAn proposed a statistical method to defend against this attack~\cite{tang2021demon}; however, it is less desirable against attacks with dynamic triggers.

\section{Conclusion}
\label{sec: Conclusion}

In this paper, we demonstrated that the existing defensive techniques heavily rely on the premise of the universal backdoor trigger, in which poisoned samples share the same trigger and thus show the same abnormal behavior. Once this prerequisite is violated, they can no longer effectively detect the advanced backdoor attacks like dynamic backdoors, where the adversary injects sample-specific triggers to each input. Based on the observation that Gramian information of dynamically trojaned data points is highly distinct from that of the benign ones, we developed Beatrix to capture not only the feature correlations but also the appropriately high-order information of the representations of benign and malicious samples, and utilized the Kernel-based two-sample testing to identify the infect labels. Experimental results show the effectiveness and robustness of our proposed approach. Beatrix can successfully defend against backdoor attacks for not only the conventional ones but also the advanced attacks, such as dynamic backdoors which can defeat the aforementioned defensive techniques.

\bibliographystyle{IEEEtran}
\bibliography{references}

\section*{Appendix}
\setcounter{section}{0}
\renewcommand{\thesection}{\Alph{section}}

\section{Detailed Discussion of Existing Backdoor Detection}

\noindent\textbf{Defending against universal backdoor.}
In universal backdoor attacks where different trojaned samples share a same trigger, the trojaned model is overfitted to the backdoor trigger~\cite{li2020rethinking}. Thus, the misclassification of a trojaned image is predominantly depended on the trigger regardless of the image content~\cite{tang2021demon, li2021invisible, nguyen2020input}.
This rudimentary setting is the Achilles heel of the existing attacks such that the overfitted trigger pattern gives rise to a series of detection methods~\cite{wang2019neural, liu2019abs, gao2019strip, chou2020sentinet, chen2019detecting}.
Based on the assumption that the backdoor trigger is sample-agnostic, existing defensive techniques can easily estimate the universal trigger~\cite{wang2019neural, liu2019abs} or detect the trojaned samples according to their common abnormal behavior~\cite{gao2019strip, chou2020sentinet, chen2019detecting}.

The success of all these defensive techniques relies on the premise of universal backdoors where the same trigger subvert all benign inputs. Once this assumption is violated, their effectiveness will be heavily vitiated.

\noindent \textbf{Defending against partial backdoor.} Firstly  highlighted in~\cite{wang2019neural}, the partial backdoor was considered as a powerful and stealthy attack since the trigger only convert samples in a specific class but has no impact on those in other classes. Although several works~\cite{wang2019neural,liu2019abs,doan2020februus} attempt to defend against the partial backdoor attacks, their main assumptions still focus on the universal backdoor attacks. Realizing the vulnerability of the existing backdoor defenses that focus on sample-agnostic (a.k.a. source-agnostic) backdoors, Tang \textit{et al.}~\cite{tang2021demon} studied the source-specific backdoor attack and showed that the trojaned and benign samples are clearly distinguishable in the feature space under the universal backdoor, but deeply tangling with each other under the partial backdoor. Therefore, They proposed Statistical Contamination Analysis (SCAn) to detect the source-specific backdoor by modeling the distributions of benign and malicious samples' representations.

\section{Theoretical  Analysis of SCAn}
\label{appendix: Theoretical  Analysis of SCAn}
SCAn models the feature distribution by a Gaussian distribution under the Linear Discriminant Analysis (LDA) assumption, \textit{i.e.}, different mean values but same covariance for the distributions of clean and trojaned feature representations. 
Formally, let the identity vectors (mean values) of clean and poisoned data be $u_{1}$ and $u_{2}$ respectively. And the covariance is denoted as $S_{\epsilon}$.  Then, the representation of a clean sample is $r_{i}^{clean} = u_{1} + e_{i}$, where $e_{i} \sim \mathcal{N}(0,S_{\epsilon})$ and $r_{i}^{clean}$ follows a Gaussian distribution  $\mathcal{N}(u_{1},S_{\epsilon})$. Similarly, the representation of a poisoned sample can be denoted as $r_{j}^{poison} = u_{2} + e_{j}$, where $e_{j} \sim \mathcal{N}(0,S_{\epsilon})$ and $r_{i}^{poison} \sim \mathcal{N}(u_{2},S_{\epsilon})$.

Let $n_1$ denote the number of clean samples and $n_2$ denotes the number of poison samples in the target (\textit{i.e.}, infected) class. Then, the mean value of the representations of all samples (clean and poison samples) in this class is:

\begin{align*}
u_{0} &= \frac{1}{N}\left(\sum^{n_1}{r_{i}^{clean}} + \sum^{n_2}{r_{j}^{poison}}\right) \\
&= \frac{1}{N}\left(\sum^{n_1}{\left(u_{1} + e_{i}\right)} + \sum^{n_2}{\left(u_{2} + e_{j}\right)}\right)\\
&= \frac{1}{N}\left(\sum^{n_1}{u_{1}} + \sum^{n_1}{e_{i}} + \sum^{n_2}{u_{2}} + \sum^{n_2}{e_{j}}\right)
\end{align*}

Recall that $e_{i}$ and $e_{j} \sim \mathcal{N}(0,S_{\epsilon})$, 
\begin{align*}
u_{0} &= \frac{n_1}{N}{u_{1}} + \frac{n_2}{N}{u_{2}}
\end{align*}

SCAn formulates the task of backdoor detection as a likelihood-ratio test problem over the feature representations of all samples (\textit{i.e.}, $R=R^{clean}\cup R^{poison}$) based on two hypotheses: 

Null Hypothesis \bm{$H$}$_{0}$: $R$ is drawn from a single normal distribution.

Alternative Hypothesis \bm{$H$}$_{1}$ : $R$ is drawn from a mixture of two normal distributions.

Therefore, the likelihood ratio under \bm{$H$}$_{1}$ hypothesis (\textit{i.e.}, the class is infected) is defined as \cite{tang2021demon}:

\begin{footnotesize}
\begin{align*}
\mathcal{L} \!\!=&\!\! \sum^{N}_{k}\left[(r_{k} \!-\! u_0)S_{\epsilon}^{-1}(r_{k} \!-\! u_0)^{T} \!\!\!-\! (r_{k} \!-\! u_m)S_{\epsilon}^{-1}(r_{k} \!-\! u_m)^{T}\right]
\end{align*}
\end{footnotesize}
\vspace{5mm}

where $m\in \left\{1,2\right\}$ is the label of the representation $r_{k}$.
\begin{footnotesize}
\begin{align}
\label{equ: likelihood ratio}
\begin{split}
\mathcal{L} \!\!=& \sum^{n1}_{i}\left[(r_{i} \!-\! u_0)S_{\epsilon}^{-1}(r_{i} \!-\! u_0)^{T} \!\!\!-\! (r_{i} \!-\! u_1)S_{\epsilon}^{-1}(r_{i} \!-\! u_1)^{T}\right] \\
&+\!\! \sum^{n2}_{j}\left[(r_{j} \!-\! u_0)S_{\epsilon}^{-1}\!(r_{j} \!-\! u_0)^{T} \!\!\!-\!\! (r_{j} \!-\! u_2)S_{\epsilon}^{-1}\!(r_{j} \!-\! u_2)^{T}\right]
\end{split}    
\end{align}
\end{footnotesize}

When $r_{k}$ is clean sample (the first term in (\ref{equ: likelihood ratio})):
\begin{footnotesize}
\begin{align*}
&\sum^{n1}_{i}\left[(r_{i} \!-\! u_0)S_{\epsilon}^{-1}(r_{i} \!-\! u_0)^{T} \!-\! (r_{i} \!-\! u_1)S_{\epsilon}^{-1}(r_{i} \!-\! u_1)^{T}\right]\\
\begin{split}
=& \sum^{n1}_{i}[(u_{1} + e_{i} - u_0)S_{\epsilon}^{-1}(u_{1} + e_{i} - u_0)^{T} \\&- (u_{1} + e_{i} - u_1)S_{\epsilon}^{-1}(u_{1} + e_{i} - u_1)^{T}]
\end{split}
\\=& \sum^{n1}_{i}\left[(u_{1} + e_{i} - u_0)S_{\epsilon}^{-1}(u_{1} + e_{i} - u_0)^{T} - e_{i}S_{\epsilon}^{-1}e_{i}^{T}\right]\\
\begin{split}
=& \sum^{n1}_{i}[(u_{1} + e_{i} - \frac{n_1}{N}{u_{1}} + \frac{n_2}{N}{u_{2}})S_{\epsilon}^{-1}(u_{1} + e_{i} - \frac{n_1}{N}{u_{1}} + \frac{n_2}{N}{u_{2}})^{T} \\&- e_{i}S_{\epsilon}^{-1}e_{i}^{T}]
\end{split}
\\=& \sum^{n1}_{i}\left[(\frac{n_2}{N}(u_{1} \!-\! u_{2}) \!+\! e_{i})S_{\epsilon}^{-1}(\frac{n_2}{N}(u_{1} \!-\! u_{2}) \!+\! e_{i})^{T} \!-\! e_{i}S_{\epsilon}^{-1}e_{i}^{T}\right]\\
\begin{split}
=& \sum^{n1}_{i}[(\frac{n_2}{N})^{2}(u_{1} - u_{2})S_{\epsilon}^{-1}(u_{1} - u_{2})^{T} \\&+ 2\frac{n_2}{N}(u_{1} - u_{2})S_{\epsilon}^{-1}e_{j}^{T} + e_{i}S_{\epsilon}^{-1}e_{i}^{T} - e_{i}S_{\epsilon}^{-1}e_{i}^{T}]
\end{split}\\
=& \sum^{n1}_{i}\left[(\frac{n_2}{N})^{2}(u_{1} \!-\! u_{2})S_{\epsilon}^{-1}(u_{1} \!-\! u_{2})^{T} \!+\! 2\frac{n_2}{N}(u_{1} \!-\! u_{2})S_{\epsilon}^{-1}e_{j}^{T}\right]
\end{align*}
Recall that $e_{i} \sim \mathcal{N}(0,S_{\epsilon})$,
\begin{align*}
=& \sum^{n1}_{i}\left[(\frac{n_2}{N})^{2}(u_{1} - u_{2})S_{\epsilon}^{-1}(u_{1} - u_{2})^{T}\right] \quad\quad\quad\quad\quad\quad\quad\quad\quad\quad\quad\quad\quad\quad\quad\quad\quad\quad
\end{align*}
Similarly, when $r_{k}$ is poison sample (the second term in (\ref{equ: likelihood ratio})):
\begin{align*}
&\sum^{n2}_{j}\left[(r_{j} \!-\! u_0)S_{\epsilon}^{-1}(r_{j} \!-\! u_0)^{T} \!-\! (r_{j} - u_2)S_{\epsilon}^{-1}(r_{j} \!-\! u_j)^{T}\right]\\
=& \sum^{n2}_{j}\left[(\frac{n_1}{N})^{2}(u_{1} - u_{2})S_{\epsilon}^{-1}(u_{1} - u_{2})^{T}\right]
\end{align*}
Therefore, the likelihood ratio of the infected class is
\begin{align}
\label{equ: scan likelihood ratio}
\nonumber \begin{split}
\mathcal{L} =& \sum^{n1}_{i}\left[(\frac{n_2}{N})^{2}(u_{1} - u_{2})S_{\epsilon}^{-1}(u_{1} - u_{2})^{T}\right] \\&+ \sum^{n2}_{j}\left[(\frac{n_1}{N})^{2}(u_{1} - u_{2})S_{\epsilon}^{-1}(u_{1} - u_{2})^{T}\right]
\end{split} \\
\nonumber =& \left[n_1(\frac{n_2}{N})^{2}+n_2(\frac{n_1}{N})^{2}\right](u_{1} - u_{2})S_{\epsilon}^{-1}(u_{1} - u_{2})^{T}\\ 
\nonumber =& \left[\frac{n_1n_2(n_1+n_2)}{N^{2}}\right](u_{1} - u_{2})S_{\epsilon}^{-1}(u_{1} - u_{2})^{T} \\ 
=& \left(\frac{n_1n_2}{N}\right)(u_{1} - u_{2})S_{\epsilon}^{-1}(u_{1} - u_{2})^{T} 
\end{align}
\end{footnotesize}

\section{Experiment Setup}
\label{sec: Experiment Setup}

\noindent \textbf{Datasets.}
To evaluate the performance of our proposed method, we take four datasets which are commonly used in backdoor-related works~\cite{tang2021demon,li2021invisible,pang2022trojanzoo,wang2019neural}:

{$\vcenter{\hbox{\tiny$\bullet$}}$} \texttt{CIFAR10}\cite{krizhevsky2009learning}. It consists of 50,000 colored training images of size 32$\times$32 and 10,000 testing images which are equally distributed on 10 classes;

{$\vcenter{\hbox{\tiny$\bullet$}}$} \texttt{GTSRB}\cite{stallkamp2012man}. It consists of 39,209 colored training images and 12,630 testing images of 43 different traffic signs. The image sizes range from 29 $\times$ 30 to 144 $\times$ 48. However, we resize them all to be 32$\times$32.

{$\vcenter{\hbox{\tiny$\bullet$}}$} \texttt{VGGFace}\cite{parkhi2015deep}. In the original dataset, there are 2,622 identities and each identity has 1,000 face images in 224$\times$224 pixels. However, about half of the links are no longer available. Following the previous work~\cite{li2021invisible}, we select the top 100 identities with the largest number of images. 
This way, we obtain 100 identities with 48,305 images. We randomly split them into training and testing samples with a ratio of 8:2. 

{$\vcenter{\hbox{\tiny$\bullet$}}$} \texttt{ImageNet}\cite{deng2009imagenet}. This dataset is related to the object classification task. Similar to~\cite{pang2020tale,pang2022trojanzoo,li2021invisible}, we randomly select a subset containing 100 classes. Out of their samples, 50,000 224$\times$224 sized images are picked for training (500 images per class) and 10,000 images are put aside for testing (100 images per class).

\noindent \textbf{Models.~} To build the classifier on CIFAR-10 and GTSRB, we use Pre-activation ResNet18~\cite{he2016identity}, following the original dynamic backdoor~\cite{nguyen2020input}. Additionally, we use VGG16~\cite{simonyan2014very} for VGGFace and ResNet101~\cite{he2016deep} for ImageNet datasets, respectively. Their top-1 accuracy on the corresponding testset is summarized in Table~\ref{tabel: information of benigh models}.

\section{Defense Performance Against ISSBA}
\label{sec: Defense Performance Against ISSBA}

In Section \ref{sec: Robustness Against Other Attacks}, we demonstrate the robustness of Beatrix against the invisible sample-specific backdoor attack (ISSBA). In this section, we provide a comprehensive comparison with the state-of-the-art defense approaches on ISSBA. In the offline setting, we generate 20 infected models with respect to the target label from 0 to 19. For online setting, we randomly select 2000 samples from each dataset as testing samples, half of which carry dynamic (or static) triggers. Similar to results on the input-aware dynamic backdoor attack, the experimental results on ISSBA demonstrate that Beatrix can significantly outperform the existing defenses, as shown in Table. \ref{tabel: comparison_on_ISSBA}.

\begin{table}[h]
\centering
\caption{\label{tabel: comparison_on_ISSBA} Defense performance against ISSBA on ImageNet.}
\resizebox{0.35\textwidth}{!}{
\begin{tabular}{ccccc}
\toprule
\multirow{2}{*}{\textbf{Scenario}}  & \multirow{2}{*}{\textbf{Method}} & \multicolumn{3}{c}{\textbf{ImageNet}} \\
\cmidrule(r){3-5}
 & & \textbf{REC}(\%) & \textbf{PRE}(\%) & \textbf{F1} (\%) \\ 
\midrule
\multirow{5}{*}{Offline} 
& NC       & 35.0  & 15.9  & 21.9    \\
& ABS      & 40.0  & 47.1 & 43.2 \\
& AC       & 90.0 & 35.3  & 50.7     \\
& SCAn     & 40.0 & 72.7 & 51.6 \\
& Beatrix     & \textbf{85.0} & \textbf{100.0} & \textbf{91.9} \\ 
\midrule
\multirow{4}{*}{Online}  
& STRIP    & 17.5 & 41.9 & 24.6     \\
& SentiNet & 0.00 & 0.00 & 0.00     \\
& SCAn     & 45.8 & 79.2 & 58.0     \\
& Beatrix     & \textbf{98.5} & \textbf{99.0} & \textbf{98.7}  \\
\bottomrule
\end{tabular}}
\end{table}

\section{Ablation Study}

We carried out a variety of ablation studies to demonstrate the effectiveness of using MAD and RMMD as the detection metrics in our approach. Herein, Beatrix-UG uses a univariate Gaussian model in the deviation measurement, and Beatrix-G assumes that the benign and malicious samples follow Gaussian distributions. In Beatrix-UG, we use the variance of Gramian features instead of MAD. That is
\begin{footnotesize}
\begin{align}
\bar{s}_j =& \  mean(\{s_{ij},\ \forall i\in\{1,2,...,|\mathcal{X}_t|\}\}), \\
variance = & \ mean(\{(s_{ij}-\bar{s}_j)^{2},\ \forall i\in\{1,2,...,|\mathcal{X}_t|\}\}).
\end{align}
\end{footnotesize}

In Beatrix-G, we alter RMMD metric to mahalanobis distance metric which assumes the data distribution is characterized by a mean and the covariance matrix (\textit{i.e.}, multivariate Gaussian distribution assumption in SCAn). Since this change only affects the identification of infected labels, we only evaluate its performance in the offline setting. 
As shown in Table~\ref{tabel: ablation_study}, for both online and offline cases, our approach with MAD and RMMD outperforms the baselines.

\begin{table}[t]
\centering
\caption{\label{tabel: ablation_study} Ablation study. The defense performance against dynamic backdoors on GTSRB and CIFAR10.}
\resizebox{0.49\textwidth}{!}{
\begin{tabular}{cccccccc}
\toprule
\multirow{2}{*}{\textbf{Scenario}}        & \multirow{2}{*}{\textbf{Method}} & \multicolumn{3}{c}{\textbf{GTSRB}} & \multicolumn{3}{c}{\textbf{CIFAR10}}  \\
\cmidrule(r){3-5}\cmidrule(r){6-8}
 & & \textbf{REC}(\%) & \textbf{PRE}(\%) & \textbf{F1} (\%) & \textbf{REC}(\%) & \textbf{PRE}(\%) & \textbf{F1}(\%) \\ 
\midrule
\multirow{3}{*}{Offline} 
& Beatrix-UG  & \textbf{97.7} & 77.8 & 86.6 & 50.0 & 83.3 & 62.5  \\
& Beatrix-G & 76.7 & 73.3 & 75.0 & 10.0 & 33.3 & 15.4  \\
& Beatrix  & 95.3 & \textbf{87.2} & \textbf{91.1} & \textbf{100.0} & \textbf{83.3} & \textbf{90.9}  \\ 
\midrule
\multirow{3}{*}{Online}  
& Beatrix-UG   & 99.5 & 93.9 & 96.6     &  95.0 & 79.0 & 86.0   \\
& Beatrix     & \textbf{99.8} & \textbf{99.8} & \textbf{99.8}     & \textbf{99.0} & \textbf{95.4} & \textbf{97.2}  \\
\bottomrule
\end{tabular}}
\end{table}

\section{Defense Performance Against Composite backdoor~\cite{lin2020composite}}
\begin{figure}[t]
\centering
\subfigure[]{
\begin{minipage}[t]{0.22\textwidth}
\centering
\includegraphics[width=0.9\linewidth]{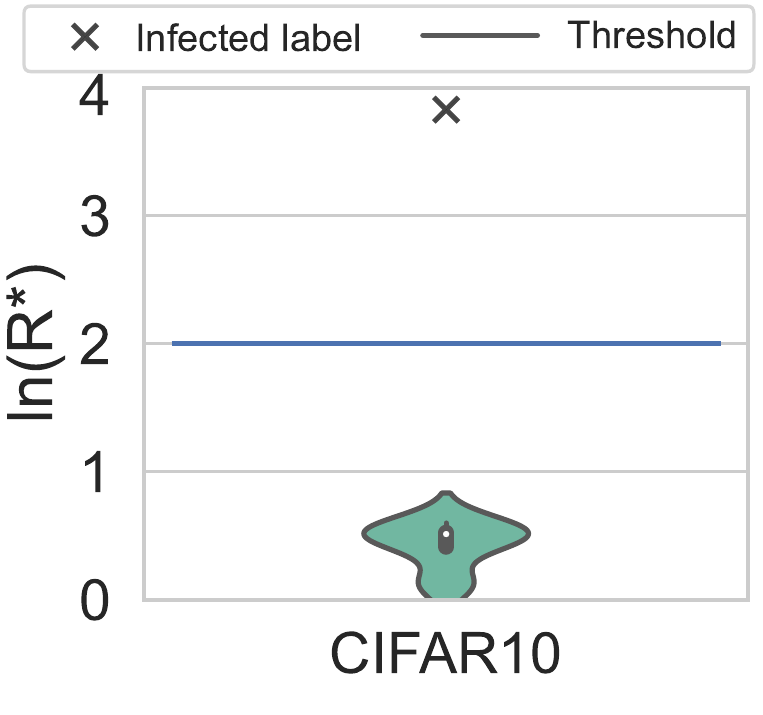}
\end{minipage}}
\subfigure[]{
\begin{minipage}[t]{0.22\textwidth}
\centering
\includegraphics[scale=0.45]{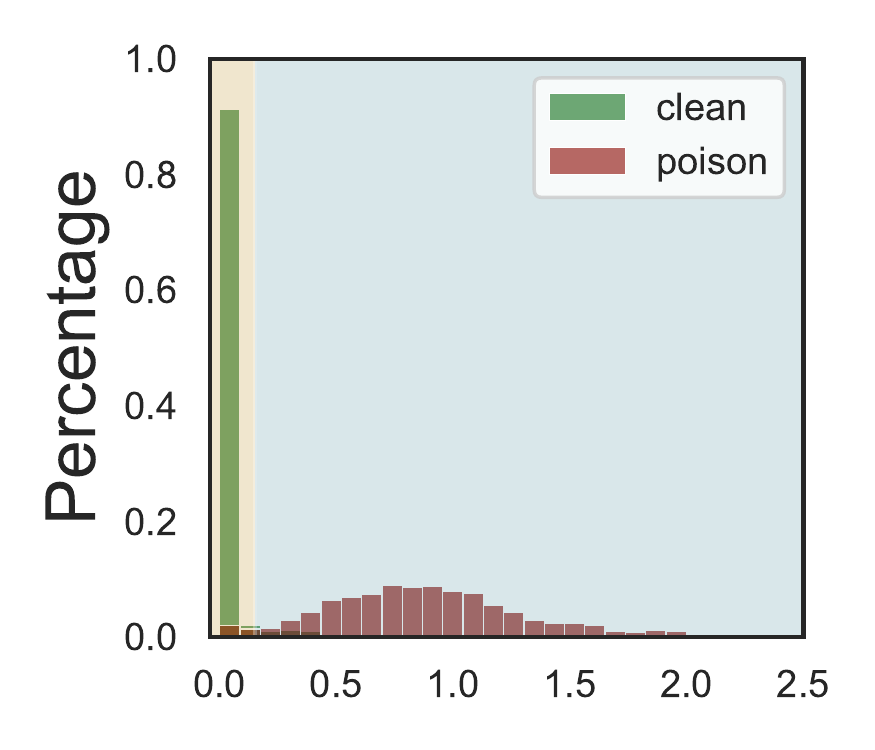}
\end{minipage}}
\caption{\label{fig: defend_against_composite} (a) The logarithmic anomaly index of infected and uninfected labels under the composite attack. (b) Deviation distribution of benign and trojaned samples in the infected class under the composite attack.}
\vspace{-1mm}
\end{figure}

Instead of injecting new features that do not belong to any output label, Lin \textit{et al.} proposed \textit{composite attack} that uses composition of existing benign features/objects as the trigger. It leverages a mixer to generate poisonous samples. The poisoned model causes targeted misclassification when the trigger composition is present. We evaluate Beatrix against this attack on the object recognition task using the code shared by the authors \cite{codecomposite}. Following the original implementation, we poison the model to predict \textit{mixer(airplane, automobile)} to \textit{bird} in CIFAR10 dataset. Figure \ref{fig: defend_against_composite} demonstrates the effectiveness of Beatrix. It shows that the anomaly index of the infected label (bird) is larger than the threshold, and those of the uninfected labels are all below the threshold. For online detection, Beatrix can effectively distinguish clean images from poisoned ones with 92.11\% TPR @ 5\% FPR and 96.88\% AUC score.

\section{Defense Performance Against \textsc{WaNet}~\cite{nguyen2021wanet}}

\begin{figure}[t]
\centering
\subfigure[]{
\begin{minipage}[t]{0.22\textwidth}
\centering
\includegraphics[width=0.9\linewidth]{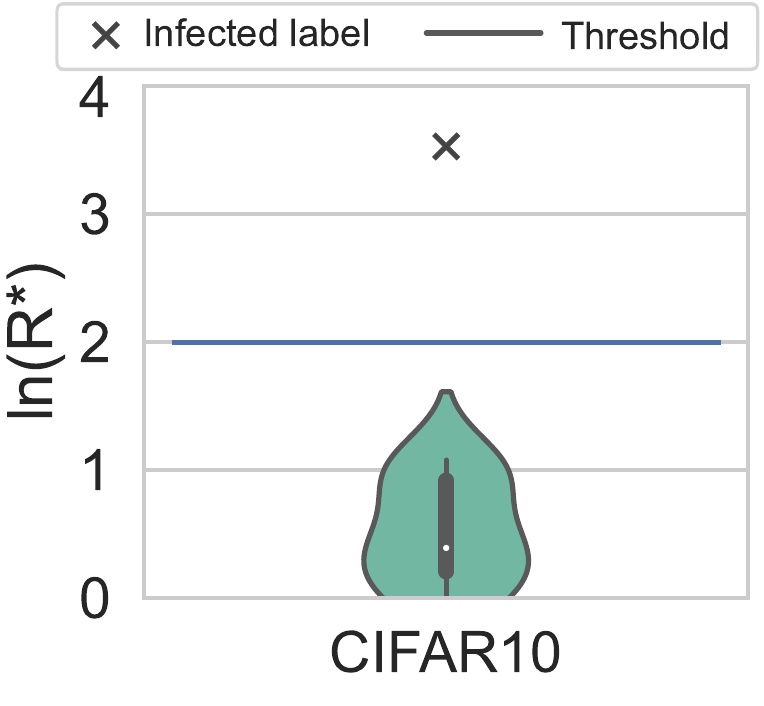}
\end{minipage}}
\subfigure[]{
\begin{minipage}[t]{0.22\textwidth}
\centering
\includegraphics[scale=0.45]{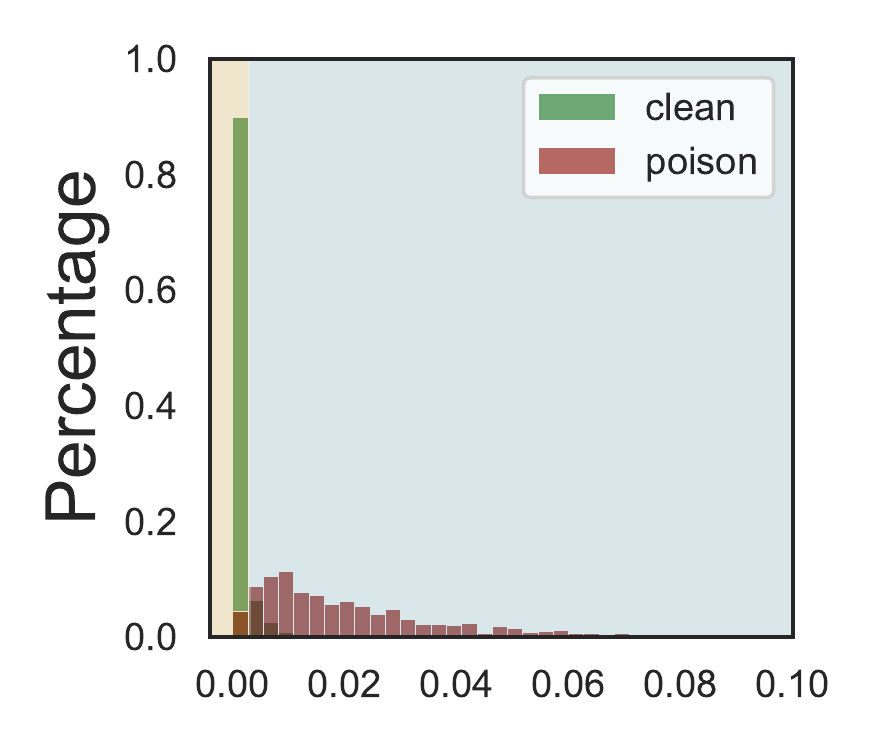}
\end{minipage}}
\caption{\label{fig: defend_against_wanet} (a) The logarithmic anomaly index of infected and uninfected labels under \textsc{WaNet}. (b) Deviation distribution of benign and trojaned samples in the infected class under \textsc{WaNet}.}
\vspace{-1mm}
\end{figure}

To improve the stealthiness of backdoor attacks, Nguyen \textit{et al.} proposed \textsc{WaNet} based on image warping. \textsc{WaNet} adopts the same elastic transformation in generating backdoor images, making the modification unnoticeable for human eyes. We put Beatrix into test against \textsc{WaNet} on CIFAR10 using the published code~\cite{codewanet}. As shown in Figure \ref{fig: defend_against_wanet}, Beatrix can effectively identify the infected label (label 0) of the poisoned model, and achieve 89.90\% TPR @ 5\% FPR and 97.94\% AUC score in online detection.

\section{Defense Performance Against Multi-trigger attack~\cite{gong2021defense}}

\begin{figure}[t]
\centering
\subfigure[]{
\begin{minipage}[t]{0.22\textwidth}
\centering
\includegraphics[scale=0.45]{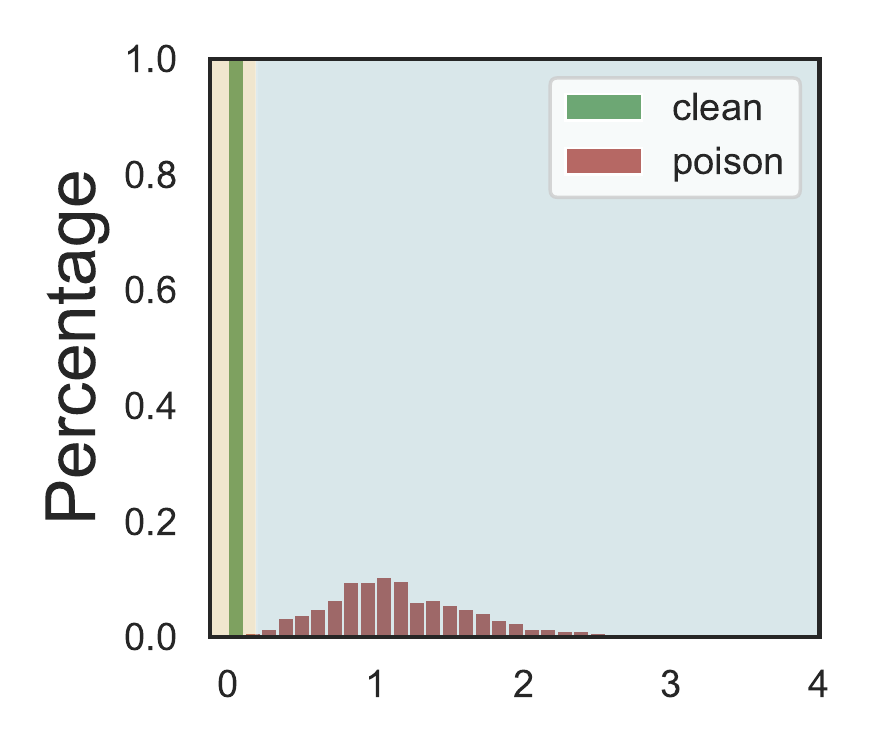}
\end{minipage}}
\subfigure[]{
\begin{minipage}[t]{0.22\textwidth}
\centering
\includegraphics[scale=0.45]{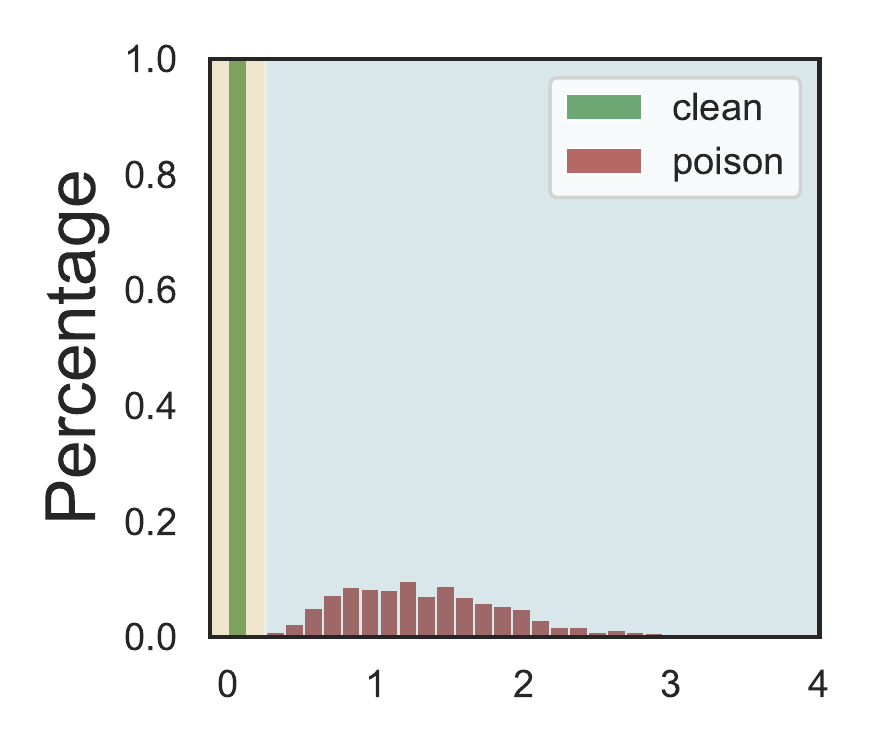}
\end{minipage}}
\subfigure[]{
\begin{minipage}[t]{0.45\textwidth}
\centering
\includegraphics[scale=0.70]{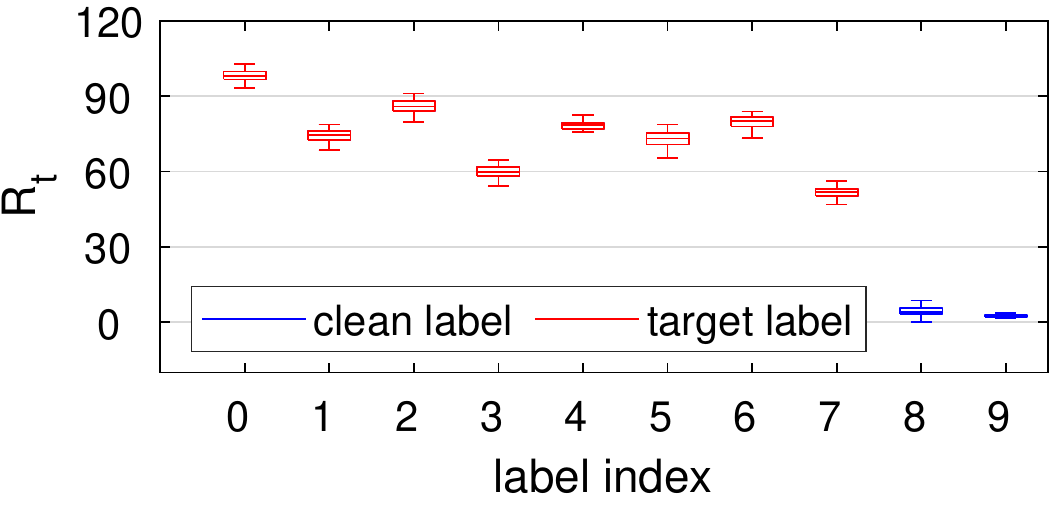}
\end{minipage}}
\caption{\label{fig: defend_against_robnet} Deviation distribution of benign and trojaned samples in the infected class under (a) the multi-trigger same-label attack and (b) the multi-trigger multi-label attack. (c) RMMD statistics $R_t$ of poisoned labels (red) and clean labels (blue) in multi-trigger multi-label attack with the number of triggers being 8.}
\vspace{-1mm}
\end{figure}

In order to increase the triggers diversity to avoid being detected, Gong \textit{et al.} proposed \textsc{RobNet} using a multi-location patching method. And they extend the attack space by designing multi-trigger backdoor attacks that can produce different triggers targeting the same or different misclassification label(s). We re-implemented this attack according to the paper \cite{gong2021defense} since the source code is not publicly available. We evaluate Beatrix against \textsc{RobNet} (\textit{i.e.}, the multi-trigger same-label attack and multi-trigger multi-label attack) on the CIFAR10 dataset. For both types of attacks, we set the number of triggers as 8 which is the maximum number of triggers used in the original paper. Beatrix can effectively distinguish the benign and trojaned samples in the infected class under both the multi-trigger same-label attack (Figure~\ref{fig: defend_against_robnet}.a) and the multi-trigger multi-label attack (Figure~\ref{fig: defend_against_robnet}.b). Similar to the all-to-all attack, the anomaly index $R_{t}^{*}$ may not be effective when more than a half of labels are infected in the multi-trigger multi-label attack. However, the RMMD statistics $R_t$ of infected labels are much larger than those of uninfected labels, which can also indicate the existence of backdoor attacks.

\section{Defense Performance on Non-Gaussian Distribution Dataset}

\begin{figure}[t]
\centering
\includegraphics[width=0.75\linewidth]{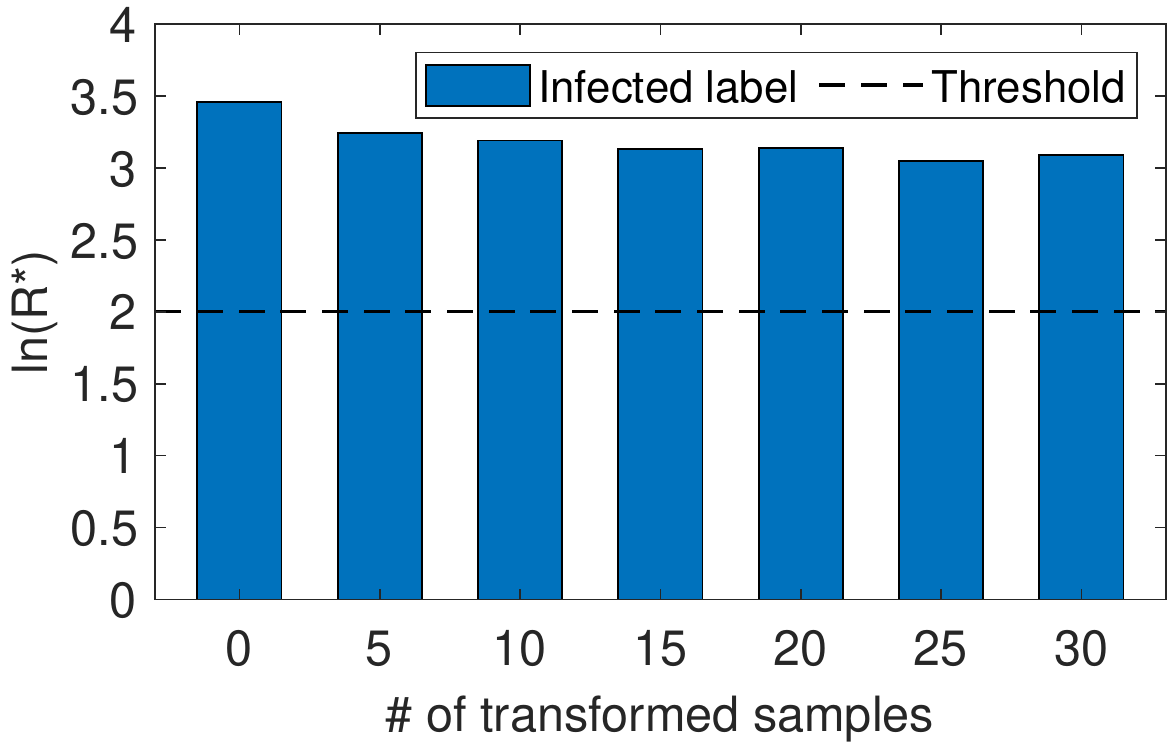}
\caption{\label{fig: defend_against_non_gaussian} The logarithmic anomaly index of infected labels when different numbers of clean images are transformed by the COLORJITTER function.}
\vspace{-1mm}
\end{figure}

We also evaluate Beatrix on a non-Gaussian distribution dataset where there are feature differences within one class. We use COLORJITTER in PyTorch to randomly change the brightness, contrast, saturation and hue of benign images (all parameters are set as 0.5). As shown in Figure~\ref{fig: defend_against_non_gaussian}, even if all 30 clean images are transformed by COLORJITTER, the computed $ln(R_t^*)$ remains effective.

\end{document}